\documentclass[aps,pre,twocolumn,superscriptaddress,longbibliography]{revtex4-1}

\pdfoutput=1

\usepackage{amsmath,amssymb}
\usepackage{bm}
\usepackage{graphicx}

\usepackage{textcomp}
\usepackage{cancel}
\usepackage{wasysym}

\usepackage{color}
\definecolor{myblue}{RGB}{65,105,225}
\definecolor{mygreen}{RGB}{34,139,34}
\definecolor{myorange}{RGB}{255,69,0}
\usepackage[colorlinks,linkcolor=myorange,urlcolor=myblue,citecolor=myblue]{hyperref}

\usepackage{mathtools}

\newcommand{\ra}{\rangle}
\newcommand{\la}{\langle}

\newcommand{\be}{\begin{equation}}
\newcommand{\ee}{\end{equation}}
\newcommand{\ben}{\begin{eqnarray}}
\newcommand{\een}{\end{eqnarray}}

\begin{document}

\title{The kinetic exclusion process: A tale of two fields}

\author{Carlos Guti\'errez-Ariza}
\email[]{carlos.gutierrez@csic.es}
\affiliation{Instituto Andaluz de Ciencias de la Tierra, CSIC--Universidad de Granada}
\affiliation{Instituto Carlos I de F\'{\i}sica Te\'orica y Computacional, Universidad de Granada, Granada 18071, Spain}

\author{Pablo I. Hurtado}
\email[]{phurtado@onsager.ugr.es}
\affiliation{Instituto Carlos I de F\'{\i}sica Te\'orica y Computacional, Universidad de Granada, Granada 18071, Spain}
\affiliation{Departamento de Electromagnetismo y F\'{\i}sica de la Materia, Universidad de Granada, Granada 18071, Spain}

\date{\today}

\begin{abstract}
We introduce a general class of stochastic lattice gas models, and derive their fluctuating hydrodynamics description in the large size limit under a local equilibrium hypothesis. The model consists in energetic particles on a lattice subject to exclusion interactions, which move and collide stochastically with energy-dependent rates. The resulting fluctuating hydrodynamics equations exhibit nonlinear coupled particle and energy transport, including particle currents due to temperature gradients (Soret effect) and energy flow due to concentration gradients (Dufour effect). The microscopic dynamical complexity is condensed in just two matrices of transport coefficients: the diffusivity matrix (or equivalently the Onsager matrix) generalizing Fick-Fourier's law, and the mobility matrix controlling current fluctuations. Both transport coefficients are coupled via a fluctuation-dissipation theorem, suggesting that the noise terms affecting the local currents have Gaussian properties. We further prove the positivity of entropy production in terms of the microscopic dynamics. The so-called kinetic exclusion process has as limiting cases two of the most paradigmatic models of nonequilibrium physics, namely the symmetric simple exclusion process of particle diffusion and the Kipnis-Marchioro-Presutti model of heat flow, making it the ideal testbed where to further develop modern theories of nonequilibrium behavior.
\end{abstract}


\maketitle

\section{Introduction}
\label{s0}

Nonequilibrium statistical physics deals with the emergent collective properties of systems composed by many degrees of freedom which are driven out of equilibrium either by external agents (as e.g. gradients, fields, etc.) or internal dissipative mechanisms. The inherent complexity of such systems makes necessary the development of simple models that, while maintaining the key ingredients to understand the phenomenon of interest, are ripped off unnecessary details which can only blur their analysis and predictive power. Chief among these simple models, stochastic lattice gases have played a pivotal role in most breakthroughs of nonequilibrium physics during the last decades \cite{spohn12a}. These include the different fluctuation theorems \cite{gallavotti95a,gallavotti95b,gallavotti95c,evans93a,evans94a,kurchan98a,lebowitz99a,jarzynski97a,jarzynski97b,hatano01a,crooks98a,crooks00a,hurtado11b}, the large deviation approach to nonequilibrium physics \cite{touchette09a,derrida07a,ellis07a} and its formulation in terms of macroscopic fluctuation theory \cite{bertini01a,bertini02a,bertini05a,bertini05b,bertini13a,bertini15a}, exact solutions distilling general features of nonequilibrium systems \cite{derrida98a,derrida98b,derrida01a,golinelli06a,blythe07a}, the discovery of new instabilities and phase transitions out of equilibrium \cite{marro05a}, or applications to biology \cite{chou11a}, active matter \cite{kourbane18a}, disordered \cite{ritort03a} and granular \cite{lasanta15a,manacorda16a,plata16a} media, soft condensed matter \cite{dunweg09a}, etc.

Despite these advances, aided by the development of stochastic lattice gases, the problem of nonequilibrium physics remains remarkably hard and open. This is due to the difficulty in combining statistics and dynamics, which always plays a main role out of equilibrium \cite{bertini15a,derrida07a} even in the relatively simpler situation of a (nonequilibrium) steady state. In equilibrium, ensemble theory teaches us how to predict the thermodynamic properties of a macroscopic system starting from its microscopic Hamiltonian, via the computation of the partition function, a sum over all possible system microscopic configurations \cite{balescu75a,pathria09a,landau13b}. Out of equilibrium there is no such a simple and general recipe in terms of \emph{configurations} due to the importance of dynamics. This lack of a general framework is a major drawback in our ability to control, manipulate and engineer systems which typically work under nonequilibrium conditions. However, in recent years, it has been shown that out of equilibrium a formally similar theoretical scheme may work at the level of \emph{trajectories} \cite{bertini15a}. The key idea now is to consider an ensemble of trajectories, characterize the probability of each possible path in phase space (either microscopically or at a mesoscopic, field-theoretic level), and from that knowledge compute the \emph{dynamical} partition function associated to any observable of interest, as e.g. the total current traversing the system or the density field \cite{bertini15a,derrida07a}. The associated dynamical free energies or \emph{large deviation functions} characterize the macroscopic behavior of the system of interest, regardless of how far from equilibrium the system is. This paradigm has proven extremely useful and predictive in simple nonequilibrium stochastic lattice gases \cite{bertini15a}, leading to a number of groundbreaking results valid arbitrarily far from equilibrium. These range from the discovery of dynamical phase transitions and spontaneous symmetry-breaking phenomena in the fluctuations of driven systems \cite{bertini05a,bodineau05a,harris05a,bertini06a,bodineau07a,lecomte07b,lecomte07c,garrahan07a,bodineau08a,garrahan09a,hedges09a,chandler10a,garrahan10a,hurtado11a,garrahan11a,pitard11a,genway12a,ates12a,speck12a,perez-espigares13a,harris13a,lesanovsky13a,hurtado14a,vaikuntanathan14a,manzano14a,jack15a,hirschberg15a,shpielberg16a,zarfaty16a,tsobgni16a,manzano16a,lazarescu17a,brandner17a,karevski17a,carollo17a,baek17a,tizon-escamilla17b,shpielberg17a,pinchaipat17a,abou18a,manzano18a,baek18a,shpielberg18a,perez-espigares18a,perez-espigares18b,chleboun18a,klymko18a,whitelam18a,vroylandt18a,rotondo18a,buca19a} to the understanding of emergent symmetries out of equilibrium \cite{hurtado11b,villavicencio14a,perez-espigares15a,kumar15a,lacoste14a,lacoste15a}, or the recent definition of universal bounds on current fluctuations in the form of thermodynamic uncertainty relations \cite{barato15a,gingrich16a,pigolotti17a,macieszczak18a}.

The exact calculation of large deviation functions from microscopic dynamics is in general a daunting task, only accomplished for a reduced number of simple low-dimensional stochastic lattice gases \cite{bertini15a,derrida07a}. Instead, for diffusive systems an alternative theoretical framework has been developed in recent years \cite{bertini15a}. In this scheme, one starts by deriving from the microscopic dynamics of the model a mesoscopic description which takes the form of a nonlinear fluctuating hydrodynamics equation (or equivalently a Langevin-type equation for the relevant locally-conserved field). This coarse-grained description is fully determined once a few transport coefficients are derived analytically from microscopics, or instead measured in experiments or simulations. A path integral formulation of the associated Langevin equation then provides the statistical weights of the possible system trajectories at this coarse-grained level, a sort of statistical physics of trajectories from which one can compute using variational methods the dynamical partition functions of interest and the associated large deviation functions \cite{bertini15a,derrida07a}. The so-called macroscopic fluctuation theory (MFT) thus allows to understand dynamic fluctuations and nonequilibrium macroscopic behavior in a broad family of driven diffusive media which goes far beyond the few exactly-solvable models mentioned above. However, the complexity of the mathematical problem has mostly restricted this program to the simpler case of models with a single locally-conserved magnitude \cite{bertini15a,derrida07a,bertini01a,bertini02a,bodineau04a,bertini05a,bertini05b,bodineau05a,tailleur07a,hurtado09a,hurtado10a,hurtado11a,krapivsky12a,meerson13a,perez-espigares13a,akkermans13a,meerson14a,meerson14c,shpielberg16a,baek17a,shpielberg17a,baek18a,shpielberg18a,garrido19a}; see however \cite{bodineau10a,bodineau11a,prados11a,cohen12a,bodineau12a,prados12a,hurtado13a,cohen14a} for the few applications of MFT to more complex scenarios. In order to systematically extend the ideas of macroscopic fluctuation theory to the more interesting case of systems characterized by several locally-conserved magnitudes coupled nonlinearly (as it is the case of e.g. realistic fluids), it is crucial to develop minimal models for this broad class of systems which, while capturing their essential ingredients (namely nonlinear coupled diffusions possibly under boundary driving), are simple enough to be amenable to both analytical calculations and extensive computer simulations. Moreover, as the starting point of MFT is the nonlinear fluctuating hydrodynamics description of the system of interest, a cornerstone in this scheme will be the passing from the microscopic stochastic dynamics to the mesoscopic, field-theoretic description of the more involved model, including the explicit computation of the relevant transport coefficients.

\begin{figure}
\includegraphics[width=9cm]{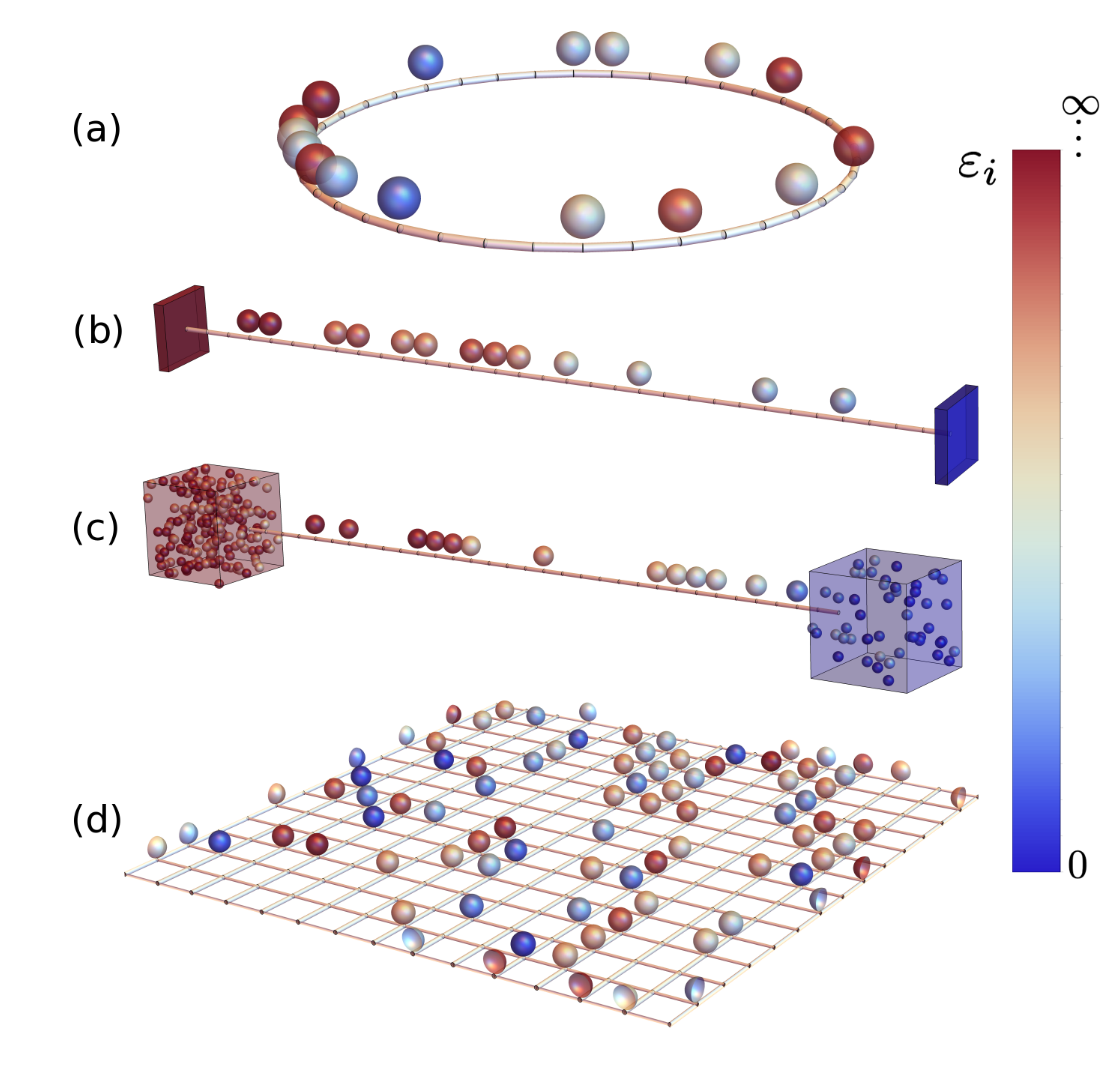}
\vspace{-1cm}
\caption{Sketch of the kinetic exclusion process (KEP), as defined on different lattices and under varying boundary conditions. Energetic particles subject to exclusion interactions jump and collide stochastically across the lattice with energy-dependent rates. Particle colors codify their energy content. Boundary conditions can be either periodic (a,d), so as to mimick an isolated gas, or open (b,c). In the latter case the system may either exchange energy with boundary thermal walls (b), or rather exchange both particles and energy with boundary reservoirs at given temperatures and chemical potentials. Boundary gradients generically drive the KEP out of equilibrium. Note that the KEP can be defined on general lattices in arbitrary dimension (d), though we restrict our discussion in this paper to $1d$ lattices for simplicity.}
\label{fig1}
\end{figure}

In this work we introduce a new model of transport, the kinetic exclusion process (KEP), characterized by the coupled nonlinear diffusion of two conserved fields, and analyze its emergent hydrodynamic behavior in the large system size limit under a local equilibrium approximation for the probability measure for microscopic configurations. The kinetic exclusion process consists in energetic particles on a general $d$-dimensional lattice subject to exclusion interactions, so each lattice site may contain at most one particle. Particles can jump stochastically to empty neighboring sites at a rate which depends (possibly in a nonlinear manner) on the particle energy content. In addition, neighboring particles can collide at a rate now dictated by the pair total energy, which is then randomly redistributed between the colliding particles. The kinetic exclusion process can be subject to different types of boundary conditions, which range from a periodic setting to study isolated dynamics to a coupling to boundary thermal walls, or even to boundary particle reservoirs at given chemical potentials and temperatures, see Fig. \ref{fig1}. For both open cases, the kinetic exclusion process can be driven out of equilibrium by introducing boundary temperature and/or chemical potential gradients. 

In the diffusive scaling limit both continuous space and time variables can be introduced, and the kinetic exclusion process can be described by two coarse-grained fields, namely the particle density $\rho(x,t)$ and a temperature field $T(x,t)$. We show below that these coupled fields evolve according to the following nonlinear fluctuating hydrodynamics equations
\ben
\partial_t \rho &=& \partial_x \left[ D_{11}(T) \partial_x \rho + D_{12}(\rho,T) \partial_x T + \xi \right] \, , \label{eq:ncont0} \\
\partial_t \left(\rho T\right) &=& \partial_x \left[ D_{21}(T) \partial_x \rho + D_{22}(\rho,T) \partial_x T + \zeta \right] \, , \label{eq:econt0} 
\een
with the diffusivity transport coefficients given by
\ben
D_{11}(T) &=& F_0(T) \, , \nonumber \label{eq:D110} \\
D_{12}(\rho,T) &=& \rho(1-\rho) F'_0(T) \, , \nonumber \label{eq:D120} \\
D_{21}(T) &=& T F_1(T) \, , \nonumber \label{eq:D210} \\
D_{22}(\rho,T) &=& \rho(1-\rho)[F_2(T)-F_1(T)] + \rho^2 F_3(T)/12 \, . \nonumber \label{eq:D220}
\een
These coefficients define the elements of the diffusivity matrix $\mathbb{D}(\rho,T)$. The functions $F_n(T)$ are a reflection of the microscopic dynamics at this mesoscopic level. They are defined as
\be
F_n(T) \equiv \int_0^\infty dy~\text{e}^{-y} y^n f(Ty) \, , \nonumber
\label{eq:F0}
\ee
with the function $f(\varepsilon)$ controlling the energy dependence of the microscopic jump/collision rates, see Eq. \eqref{eq:CollisionRate} below. Note that these functions $F_n(T)$ obey a recurrence relation $F_{n+1}(T) = (n+1) F_n(T) + T F'_n(T)$, with $'$ denoting derivative with respect to the argument. The fields $\xi(x,t)$ and $\zeta(x,t)$ in Eqs. \eqref{eq:ncont0}-\eqref{eq:econt0} are two Gaussian white noises with zero mean and correlators
\ben
\la \xi(x,t) \xi(x't')\ra &=& \frac{1}{L} M_{11}(\rho,T) \delta(x-x') \delta(t-t') \, , \nonumber \\
\la \xi(x,t) \zeta(x't')\ra &=& \frac{1}{L} M_{12}(\rho,T) \delta(x-x') \delta(t-t') \, , \label{eq:corrfields20} \\
\la \zeta(x,t) \zeta(x't')\ra &=& \frac{1}{L} M_{22}(\rho,T) \delta(x-x') \delta(t-t') \, , \nonumber 
\een
with $L$ the system size, so these noise terms are weak in the large $L$ limit. Such weak gaussian fluctuations are expected to emerge for most situations of interest in the appropriate mesoscopic limit as a result of a central limit theorem: even though microscopic interactions can be highly complicated, the ensuing fluctuations affecting the slow (hydrodynamic) degrees of freedom typically result from a large superposition of random events at the microscale which leads to Gaussian statistics at the mesoscale. We further show that the transport coefficients appearing in Eq. \eqref{eq:corrfields20} are 
\ben
M_{11}(\rho,T) &=& 2 \rho(1-\rho) F_0(T) \, , \nonumber \\
M_{12}(\rho,T) &=& 2 \rho(1-\rho) T F_1(T) = M_{21}(\rho,T) \, , \nonumber \label{eq:mobil0} \\
M_{22}(\rho,T) &=& 2 \rho(1-\rho) T^2 F_2(T) + \rho^2 T^2 F_3(T)/6 \, . \nonumber
\een
These transport coefficients define the elements of the \emph{symmetric} mobility matrix $\mathbb{M}(\rho,T)$ which controls the coupled fluctuations of the particle and energy current fields in the KEP. Interestingly, the mobility matrix $\mathbb{M}(\rho,T)$ can be simply related to the Onsager's matrix $\mathbb{L}(\rho,T)$ of phenomenological transport coefficients relating dissipative fluxes and thermodynamic forces \cite{de-groot13a}, namely
\be
\mathbb{M}(\rho,T) = 2 \, \mathbb{L}(\rho,T) \, .
\label{eq:FDT}
\ee
This is just an expression of the general fluctuation-dissipation theorem linking thermal fluctuations and the response to a small driving in microreversible systems (i.e. systems obeying detailed balance) \cite{de-groot13a}, which supports the Gaussian character of the noise terms affecting the local currents. We identify the different thermodynamic forces for the KEP, which allows to relate the Onsager's matrix $\mathbb{L}(\rho,T)$ with the diffusivity matrix $\mathbb{D}(\rho,T)$, thus leading to a direct relation between $\mathbb{D}(\rho,T)$ and $\mathbb{M}(\rho,T)$. It is important to note that the transport coefficients entering the diffusivity (equivalently Onsager's) and mobility matrices depend nonlinearly on the local particle density and temperature fields and on the function $f(\varepsilon)$ controlling the energy dependence of the microscopic jump/collision rate, see below. Note also that the fluctuating hydrodynamic equations \eqref{eq:ncont0}-\eqref{eq:econt0} capture the possibility of (i) a particle flow in the absence of a density gradient, due exclusively to a temperature gradient (Soret effect), and (ii) an energy current due exclusively to the presence of a density gradient (Dufour effect) \cite{de-groot13a}.

The kinetic exclusion process has the additional interest of converging in two different limits to two key models of nonequilibrium physics, which underlie many of the most exciting recent discoveries in nonequilibrium statistical physics. These two limiting models are the simple symmetric exclusion process (SSEP) \cite{spitzer70a,derrida98a,derrida98b,golinelli06a} on one hand, a cornerstone in the understanding of the physics of diffusion, and the Kipnis-Marchioro-Presutti (KMP) model of heat conduction \cite{kipnis82a}, a key model in transport theory. Groundbreaking exact results have been obtained for both models, including the first rigorous derivation of the elusive Fourier's law of heat conduction from microscopic dynamics in the KMP model \cite{kipnis82a}, and one of the very few exact determinations of the steady state probability measure in a nonequilibrium system \cite{derrida98a,derrida98b,derrida01a,golinelli06a}. The KEP model introduced here, having these two canonical models as well-defined limits, opens up interesting new avenues of research in the field of exactly-solvable nonequilibrium models. In particular, it is well-known that both models can be mapped onto integrable quantum spin systems \cite{derrida98a,giardina07a,giardina09a}, a mapping which has proven crucial to obtain exact solutions in both cases. This suggests to extend these spin mappings to the more general KEP model, with the aim of gaining insights on its physics by using techniques developed for integrable quantum systems. In addition, the microscopic dynamics of the KEP introduces nonlinear couplings between the different conserved fields, broadening the possible spectrum of applications and allowing for a direct contact with realistic systems where nonlinear interactions are the rule. In particular, the KEP is an ideal lab to model the physics of compressible quiescent fluids \cite{landau13a}, and can be trivially generalized to arbitrary dimension.

We note here that there have been several other attempts in literature to construct fluctuating hydrodynamics with two (or even three or more) locally-conserved fields starting from microscopics, for instance by considering systems with local conservation of mass and momentum (rather than energy), see e.g. \cite{spohn12a,mendl13a,spohn14a,das14a,basile14a,spohn15a,lasanta15a,komorowski16a,plata16a,manacorda16a,olla19a} and references therein; see also \cite{castro-alvaredo16a,bertini16a,eisert15a} for integrable systems with many conservation laws. The reader may also ponder the interest of the fluctuating hydrodynamics derived in this paper at the light of the celebrated Landau-Lifshitz classical fluctuating hydrodynamics, well-established theoretically and tested in detailed experiments based on light and neutron scattering methods \cite{landau13a}. First, it is important to note that Landau-Lifshitz theory is a phenomenological framework developed to deal with fluctuations in fluids in \emph{thermodynamic equilibrium states}. As such, Landau-Lifshitz theory is a linear theory where transport coefficients are constant, material-dependent magnitudes. On the other hand, nonlinear fluctuating hydrodynamics  theories as the one derived here (with transport coefficients depending --usually in a nonlinear manner-- on the local values of the hydrodynamic fields) are essential to capture most interesting nonequilibrium effects. In particular, this nonlinear dependence on local fields is crucial to describe fluctuations in fluids in nonequilibrium steady state and the associated universal fat tails and long-range correlations \cite{zarate06a}. Moreover, the dependence of transport coefficients on the local hydrodynamic fields is also important to understand the large-deviation statistics of fluids under nonequilibrium conditions within macroscopic fluctuation theory \cite{bertini15a}. In this sense, nonlinear fluctuating hydrodynamics supersedes (and includes) the original work of Landau and Lifshitz, providing a broader framework to understand nonequilibrium behavior.

We structure this paper as follows. In Section \S\ref{s2} we introduce the kinetic exclusion process in detail, paying special attention to the definition of the jump and collision rates and their possible dependence on the local energy content. We also discuss the two different limiting cases mentioned above, i.e. the SSEP and the KMP models. Section \S\ref{s3} is devoted to the derivation of the coupled hydrodynamic equations for the two relevant conserved fields in terms of the microscopic dynamics. For that purpose, we first derive the microscopic balance equation for the density and energy fields in Section \S\ref{s3a}. Using a local equilibrium approximation for the probability measure described in Section \S\ref{s3b}, we obtain the constitutive relations for the particle and energy currents in Section \S\ref{s3c}, leading to the full nonlinear hydrodynamics equations. We then analyze the entropy production in the kinetic exclusion process in Section \S\ref{s3d}, identifying the correct thermodynamic forces and obtaining the associated Onsager's matrix in terms of the diffusivity matrix. We also prove here the positivity of entropy production in terms of the microscopic dynamics. Section \S\ref{s4} studies the hydrodynamic fluctuations which affect both particle and energy currents. We derive the nonlinear amplitudes of the noise terms affecting both current fields, as well as their cross-correlations, and show that these noises are weak, ${\cal O}(L^{-1})$, in the large system size limit. Our results explicitly show the connection between the mobility matrix measuring the amplitude of current fluctuations and the Onsager's matrix, proving a fluctuation dissipation theorem and hinting at the Gaussian character of these noise terms. A summary of the main results of the paper, together with a physical discussion thereof, is given in Section \S\ref{s5}. Finally, the appendices deal with some technical details that, for the sake of clarity, we have preferred to omit in the main text.

\section{The kinetic exclusion process}
\label{s2}

The kinetic exclusion process is an continuous-time Markovian interacting particle system defined on a lattice, possibly driven by boundary gradients and characterized by particle exclusion and random energy exchanges, which lead to nonlinear coupled mass and energy diffusion. For simplicity the results in this paper are restricted to one-dimensional ($1d$) lattices, though they can be easily extended to arbitrary dimension. 

We focus first on periodic boundaries for simplicity, we discuss below coupling to boundary reservoirs. The model is thus defined on a $1d$ periodic lattice with $L$ sites where $N\le L$ particles evolve in time. Particles are subject to exclusion interactions, so no two particles can overlap in the same position. Each lattice site $k\in[1,L]$ is hence characterized by an occupation number $n_{k,s}=0,1$ at step $s$ of the dynamics. In addition, each particle has an \emph{energy} which determines how it moves in the lattice and collides with neighboring particles. In this way we can associate an energy $\varepsilon_{k,s}\in\mathbb{R}_0^+$ to an occupied site $k$, while empty lattice sites have zero energy. A microscopic configuration at a given time step $s$ is hence given by $\bm{\nu}_s=\{(n_{k,s},\varepsilon_{k,s}),k=1,\ldots,L\}$. The dynamics is stochastic and Markovian, and proceeds in continuous time as follows. In an elementary step a pair of nearest neighbors sites $(k,k+1)$, identified by the index of its leftmost site $k$, is randomly chosen with a probability
\begin{equation}
P(k|\bm{\nu}_s)=\frac{\Theta(n_{k,s}+n_{k+1,s}) ~f\left( \varepsilon_{k,s}+\varepsilon_{k+1,s}\right)}{L~\Omega_L(\bm{\nu}_s)}\, ,
\label{eq:CollisionRate}
\end{equation}
where $\Theta(n)\equiv \min(1,n)$ is 0 if there are no particles at pair $(k,k+1)$ and 1 otherwise, and $f(\varepsilon)\ge 0$ is a function that captures the energy dependence of the particle stochastic motion. A physically sound choice for this function is $f(\varepsilon)=\varepsilon^\beta$ with $\beta\in\mathbb{R}^+_0$ \cite{gutierrez-ariza19a}, though other choices are also possible. As soon as $f$ is an increasing function of the local energy, particles with higher energy will jump and collide more often, leading to a nonlinear and realistic density-energy coupling \cite{de-groot13a}. The numerator in Eq. (\ref{eq:CollisionRate}) defines the transition rate of the continuous-time stochastic dynamics, which obeys detailed balance with respect to the equilibrium measure, while the normalization factor in Eq. (\ref{eq:CollisionRate}) is
\be
\Omega_L(\bm{\nu}_s)= \frac{1}{L} \sum_{k'=1}^{L}\Theta(n_{k',s}+n_{k'+1,s})~f\left(\varepsilon_{k',s}+\varepsilon_{k'+1,s}\right)  \, ,
\label{eq:CollisionRate2}
\ee
with the identification of site $k=L+1$ with $k=1$ (periodic boundary conditions). The $L$-factor in the denominator has been included to make $\Omega_L(\bm{\nu}_s)$ finite in the large system size limit $L\to\infty$. The chosen nearest neighbors pair $(k,k+1)$ must contain at least one particle, as empty pairs $(\fullmoon\fullmoon)$ cannot be selected due to the condition $\Theta(0)=0$. If the chosen pair contains one particle, i.e. either $(\newmoon\fullmoon)$ or $(\fullmoon\newmoon)$, the particle jumps to the neighboring empty site and the local variables are exchanged, i.e. $n_{k,s+1}=n_{k+1,s}$ and $n_{k+1,s+1}=n_{k,s}$ for the occupation numbers, and $\varepsilon_{k,s+1}=\varepsilon_{k+1,s}$ and $\varepsilon_{k+1,s+1}=\varepsilon_{k,s}$ for the energies. On the other hand, if the chosen pair contains two particles $(\newmoon\newmoon)$, they \emph{collide} by randomly exchanging their energy so that the total energy of the pair is conserved, i.e. $n_{k,s+1}=n_{k,s}$ and $n_{k+1,s+1}=n_{k+1,s}$ and
\ben
\varepsilon_{k,s+1} &=& \alpha(\varepsilon_{k,s}+\varepsilon_{k+1,s}) \, , \label{eq:coll} \\
\varepsilon_{k+1,s+1} &=& (1-\alpha)(\varepsilon_{k,s}+\varepsilon_{k+1,s}) \, , \nonumber
\een
with $\alpha\in[0,1]$ a random number with homogeneous probability density function (pdf) $P(\alpha)=1$. Thus $\alpha$ represents the fraction of the pair total energy which remains at site $k$ after the collision. This stochastic energy redistribution mechanism, originally proposed in \cite{kipnis82a}, has been generalized to a number of mass transport models \cite{evans04a,evans05a,evans06a}. In any case, after the jump/collision step, time is incremented by a random interval, i.e. $\tau_{s+1}=\tau_{s} + \delta \tau_s/L$, with $\delta \tau_s$ drawn from a Poisson distribution $P(\delta \tau_s)=\Omega_L(\bm{\nu}_s)^{-1}\exp[- \delta \tau_s/\Omega_L(\bm{\nu}_s)]$, so the typical value of the time increment $\delta \tau_s/L$ equals the inverse of the total exit rate from the current microscopic configuration, $[L~\Omega_L(\bm{\nu}_s)]^{-1}$.

In case of open boundary conditions, the boundary sites $k=1,~L$ interact with external reservoirs located at $k=0,~L+1$, respectively. These reservoirs can exchange both energy and particles with the system (macrocanonical case) or only energy (canonical case). We focus now on the more general macrocanonical case, the canonical one being very similar. The reservoir sites $k=0,~L+1$ are therefore characterized by a chemical potential $\mu_{l,r}$ and a temperature $T_{l,r}\ge 0$ for the left ($l$) and right ($r$) boundary, respectively. To mimic the coupling with the reservoirs, a particle with energy $\varepsilon$ is injected on site $1$ with a rate $\beta_l(\varepsilon)$ if the site is empty, or removed from site $1$ with a rate $\gamma_l(\varepsilon)$ if the site is occupied (in a similar manner, for site $L$ particles with energy $\varepsilon$ are injected and removed with rates $\beta_r(\varepsilon)$ and $\gamma_r(\varepsilon)$, respectively). To ensure the time-reversivility of the microscopic dynamics, the previous boundary transition rates must fulfill the local detailed balance condition \cite{bodineau07a}. In particular, if $j_p=+1~ (-1)$ and $q_p=+\varepsilon~ (-\varepsilon)$ are the particle and energy currents flowing from reservoir $p=l,r$ to the system in an injection (removal) event, the local detailed balance condition for the boundary rates then reads \cite{bodineau07a}
\be
z_p^{-j_p} \, \text{e}^{q_p/T_p} \beta_p(\varepsilon) = \gamma_p (\varepsilon) \, ,
\label{eq:detbal}
\ee
where $z_p\equiv \exp(\mu_p/T_p)$ is the fugacity of reservoir $p=l,r$ with chemical potential $\mu_{l,r}$. This leads to the condition $\beta_p(\varepsilon)/\gamma_p (\varepsilon)=z_p \exp(-\varepsilon/T_p)$. A possible choice for the injection and removal rates at reservoir $p=l,r$ is then
\be
\beta_p(\varepsilon) = \beta_p^{0} \text{e}^{-\varepsilon/T_p} \, , \qquad \gamma_p (\varepsilon) = \gamma_p^{0} \, ,
\label{eq:ratesb}
\ee
with $\beta_p^{0}$ and $\gamma_p^{0}$ constants which set the overall timescale for boundary updates. These rates then correspond to reservoirs with densities
\be
\rho_p = \frac{T_p \beta_p^{0}}{\gamma_p^{0}+T_p \beta_p^{0}} \,
\label{eq:ratesb2}
\ee
see Eq. \eqref{eq:mu} below relating the chemical potential, temperature and density for the KEP. These simple rules can be easily modified to take into account thermal boundaries allowing only energy exchanges, and in any case both the transition probability \eqref{eq:CollisionRate} and the normalization factor \eqref{eq:CollisionRate2} must be suitably modified with the associated boundary terms. Note that for the open case, whenever the boundary temperatures and/or fugacities are different, $T_l\ne T_r$ or $z_l\ne z_r$, the existing gradient typically drives the system to a nonequilibrium steady state characterized by non-vanishing currents \cite{de-groot13a}.

The kinetic exclusion process so defined leads to coupled nonlinear energy and mass transport, as described by a non-trivial matrix of transport coefficients controlling its hydrodynamic behavior that we will characterize below. Moreover, as mentioned in the introduction, the KEP contains as limiting cases some of the most paradigmatic models of transport in nonequilibrium physics, cornerstones in most of the recent breakthroughs in this field, turning the KEP into an ideal ground for further advances. For instance, filling the KEP lattice with the maximum number of particles allowed (i.e. $N=L$) and setting $\rho_l=1=\rho_r$ (in case of open boundaries), no particle motion is possible due to the exclusion constraint. In this way the only dynamic observable in the system is the energy of the particles, which diffuses across the chain in contact with two boundary reservoirs possibly at different temperatures. In the simplest case of homogeneous, energy-independent collision rate, i.e. $f(\varepsilon)=1$ in Eq. (\ref{eq:CollisionRate}) above, we recover the standard Kipnis-Marchioro-Presutti (KMP) model of heat conduction \cite{kipnis82a}, while energy-dependent rates lead to a recently introduced nonlinear version of this model \cite{hurtado12a} which can be modified to include dissipation \cite{prados11a,prados12a,hurtado13a,lasanta16a}. The KMP model of heat transport is a pillar in nonequilibrium statistical physics which has been used as a benchmark to prove rigorous results (as e.g. the first microscopic exact derivation of Fourier's law \cite{kipnis82a}) and to test theoretical advances, ranging from the additivity principle for current fluctuations \cite{bodineau04a,hurtado09c,hurtado10a} or the Gallavotti-Cohen fluctuation theorem \cite{gallavotti95a} and its generalizations \cite{hurtado11b}, to the existence of spontaneous symmetry-breaking transitions at the fluctuating level \cite{hurtado11a}. Similarly, a KEP model with arbitrary density and boundary gradients but again with energy-independent jump rates ($f(\varepsilon)=1$) reduces to the symmetric simple exclusion process (SSEP) of particle diffusion \cite{derrida98a} as far as particle degrees of freedom are concerned. This model, considered as the prototypical stochastic model for diffusive phenomena and a paradigm in nonequilibrium behavior, has been instrumental in the discovery of exact results arbitrarily far from equilibrium, Bethe ansatz approximations, the understanding of nonequilibrium phase transitions, growth processes and rough interfaces, etc. \cite{derrida98a,golinelli06a,chou11a} These interesting limiting cases, whose algebraic structure can be understood from their mapping to integrable quantum spin models \cite{derrida98a,giardina07a,giardina09a}, together with the non-trivial energy-density coupling which emerges under general jump/collision rules, make the KEP an ideal lab to further advance in our understanding of nonequilibrium phenomena and their characterization using tools from fluctuating hydrodynamics \cite{zarate06a} and macroscopic fluctuation theory \cite{bertini15a,derrida07a,hurtado14a}.

\section{Hydrodynamics for two conserved fields}
\label{s3}

\subsection{Microscopic balance equations}
\label{s3a}

The dynamics defined for the kinetic exclusion process in the previous section can be written now in algebraic terms. With the aim of deriving the model bulk hydrodynamic behavior, we focus now for simplicity on the periodic boundary conditions case (open boundaries can be taken into account later on via boundary conditions for the hydrodynamic fields). Let $\eta=\{\bm{\nu}_0,\bm{\nu}_1,\ldots\}$ define a particular trajectory of the stochastic process. For a given initial state $\bm{\nu}_0$, this trajectory is fully defined by a sequence of pairs $\{(k_s,\alpha_s), s=1,\ldots\}$, with $k_s$ identifying the pair of nearest neighbors randomly drawn from the pdf \eqref{eq:CollisionRate} where the jump/collision event happens at step $s$, and $\alpha_s\in[0,1]$ being a homogeneous random number to be used for the random energy exchange in case of a collision in pair $k_s$. For a given trajectory $\eta$, the occupation number of an arbitrary site $i\in[1,L]$ at step $s+1$ can be written as
\ben
n_{i,s+1}^{(\eta)} &=&  n_{i,s}^{(\eta)} \left(1 - \delta_{k_s,i} - \delta_{k_s-1,i} \right) \label{eq:nevol} \\
&+& n_{i,s}^{(\eta)} n_{i+1,s}^{(\eta)} \delta_{k_s,i} + n_{i-1,s}^{(\eta)} n_{i,s}^{(\eta)} \delta_{k_s-1,i} \nonumber \\
&+& (1-n_{i,s}^{(\eta)}) ~n_{i+1,s}^{(\eta)} \delta_{k_s,i} + (1-n_{i,s}^{(\eta)}) ~n_{i-1,s}^{(\eta)} \delta_{k_s-1,i} \, . \nonumber
\een
The first line in the previous equation accounts for the possibility that the occupation number of site $i$ remains unchanged at step $s$ because $k_s\ne i,i-1$. The next line accounts for the possibility of a collision of a particle at site $i$ with another particle at a neighboring site $i\pm 1$, while the last line describes a possible jump from (to) site $i$ to (from) a neighboring site. Equivalently, and following the same reasoning, the energy of site $i$ at step $s+1$ is
\begin{table}
\begin{center}
\begin{tabular}{|c|c|c|c|c|}
\hline 
 & $\newmoon\newmoon$ & $\newmoon\fullmoon$ & $\fullmoon\newmoon$ & $\fullmoon\fullmoon$\tabularnewline
\hline 
\hline 
$j_{i,s}^{(\eta)}$ & $0$ & $+1$ & $-1$ & $0$\tabularnewline
\hline 
$q_{i,s}^{(\eta)}$ & $\varepsilon_{i,s}^{(\eta)}-\alpha_s\left(\varepsilon_{i,s}^{(\eta)}+\varepsilon_{i+1,s}^{(\eta)}\right)$ & $+\varepsilon_{i,s}^{(\eta)}$ & $-\varepsilon_{i+1,s}^{(\eta)}$ & $0$\tabularnewline
\hline 
\end{tabular}
\caption{Particle and energy currents involved in an elementary collision/jump event given the initial state.}
\label{table1}
\end{center}
\end{table}
\ben
\varepsilon_{i,s+1}^{(\eta)} &=&  \varepsilon_{i,s}^{(\eta)} \left(1 - \delta_{k_s,i} - \delta_{k_s-1,i} \right) \label{eq:eevol} \\
&+& \alpha_s(\varepsilon_{i,s}^{(\eta)} + \varepsilon_{i+1,s}^{(\eta)})~n_{i,s}^{(\eta)} n_{i+1,s}^{(\eta)} \delta_{k_s,i} \nonumber \\
&+& (1-\alpha_s)(\varepsilon_{i,s}^{(\eta)} + \varepsilon_{i-1,s}^{(\eta)})~n_{i-1,s}^{(\eta)} n_{i,s}^{(\eta)} \delta_{k_s-1,i} \nonumber \\
&+& \varepsilon_{i+1,s}^{(\eta)} (1-n_{i,s}^{(\eta)}) ~n_{i+1,s}^{(\eta)} \delta_{k_s,i} \nonumber \\
&+& \varepsilon_{i-1,s}^{(\eta)} (1-n_{i,s}^{(\eta)}) ~n_{i-1,s}^{(\eta)} \delta_{k_s-1,i} \, , \nonumber
\een
These expressions can be simplified by grouping together the terms affecting the same pair. In this way, the evolution equation for the occupation number simplifies to
\be
n_{i,s+1}^{(\eta)} - n_{i,s}^{(\eta)} = \delta_{k_s,i-1} \left(n_{i-1,s}^{(\eta)} - n_{i,s}^{(\eta)} \right) - \delta_{k_s,i} \left(n_{i,s}^{(\eta)} - n_{i+1,s}^{(\eta)} \right) \label{eq:nevol2}
\ee
or equivalently
\be
n_{i,s+1}^{(\eta)} - n_{i,s}^{(\eta)} = j_{i-1,s}^{(\eta)} - j_{i,s}^{(\eta)} \, ,
\label{eq:nevol3}
\ee
where we have defined the particle current across a pair $(i,i+1)$ at step $s$ as
\be
j_{i,s}^{(\eta)} \equiv \delta_{k_s,i} \left(n_{i,s}^{(\eta)} - n_{i+1,s}^{(\eta)} \right)\, .
\label{eq:currn}
\ee
One can easily check that this definition yields a particle current value of $+1$ whenever a particle jumps to the right ($i\to i+1$), a value $-1$ when a particle jumps to the left ($i+1\to i$), and 0 otherwise. Eq. (\ref{eq:nevol3}) is a microscopic balance equation for the particle number which expresses the local conservation law for this observable. Proceeding now in an equivalent manner for the local energy, we obtain another microscopic balance equation
\be
\varepsilon_{i,s+1}^{(\eta)} - \varepsilon_{i,s}^{(\eta)} = q_{i-1,s}^{(\eta)} - q_{i,s}^{(\eta)} \, ,
\label{eq:eevol2}
\ee
with the energy current $q_{i,s}^{(\eta)}$ defined as the energy flowing to the right across a given pair, i.e.
\ben
q_{i,s}^{(\eta)} &\equiv&  \delta_{k_s,i} \left[n_{i,s}^{(\eta)} \varepsilon_{i,s}^{(\eta)} - n_{i+1,s}^{(\eta)} \varepsilon_{i+1,s}^{(\eta)} \right. \nonumber \\
&+& \left. \left((1-\alpha_s) \varepsilon_{i+1,s}^{(\eta)} - \alpha_s \varepsilon_{i,s}^{(\eta)} \right)n_{i,s}^{(\eta)} n_{i+1,s}^{(\eta)} \right] \, .
\label{eq:curre}
\een
Table \ref{table1} summarizes for concreteness the particle and energy currents involved in each type of elementary jump/collision step defined by the initial pair state. 

The hydrodynamic or average evolution equations for the local particle and energy densities can be now obtained by averaging over all possible trajectories $\eta$ of the stochastic process, weighted by their corresponding probability. This procedure is then equivalent to averaging over all possible sequences of pairs $\{(k_s,\alpha_s), s=1,\ldots\}$ of independent random numbers $k_s$ and $\alpha_s$, each with its probability distribution, and over all initial states, weighted by some initial distribution. Averaging in this way the microscopic balance equations (\ref{eq:nevol3}) and (\ref{eq:eevol2}) following Eq. (\ref{eq:ave}) leads to
\ben
\la n_{i}\ra_{s+1} - \la n_{i}\ra_s &=& \la j_{i-1} \ra_s - \la j_{i}\ra_s \, , \label{eq:aven} \\
\la \varepsilon_{i}\ra_{s+1} - \la \varepsilon_{i}\ra_s &=& \la q_{i-1} \ra_s - \la q_{i}\ra_s \, , \label{eq:avee}
\een
where we have defined the average $\la A_i\ra_{s}$ of an arbitrary local observable $A_i(\bm{\nu};k_s,\alpha_s)$ associated to pair $i=(i,i+1)$ as
\be
\la A_i\ra_{s} = \sum_{\bm{\nu},k_s,\alpha_s} A_i(\bm{\nu};k_s,\alpha_s)~P(k_s|\bm{\nu})~P(\bm{\nu};s) \, ,  \label{eq:ave}
\ee
where $P(\bm{\nu};s)$ is the probability of finding the system in configuration $\bm{\nu}$ at step $s$, and we have already used that $P(\alpha_s)=1$. Note that we have simplified our notation by dropping the trajectory superindex $\eta$ and the time step subindex $s$ in the state variables. For configurational observables, as e.g. $A_i(\bm{\nu})=n_i$ or $\varepsilon_i$, we simply have that $\la A_i\ra_{s}= \sum_{\bm{\nu}} A_i(\bm{\nu})~P(\bm{\nu};s)$ due to the normalization of $P(k_s|\bm{\nu})$. On the other hand, from the definitions (\ref{eq:currn}) and (\ref{eq:curre}) above, the average currents can be written as
\begin{widetext}
\ben
\la j_{i}\ra_s &=& \frac{1}{L} \left\la \frac{(n_i-n_{i+1})~f(\varepsilon_i+\varepsilon_{i+1})}{\Omega_L(\bm{\nu})}\right\ra_s \, , \label{eq:currn2} \\
\la q_{i}\ra_s &=& \frac{1}{L} \left\la \frac{\left[n_i\varepsilon_i -n_{i+1}\varepsilon_{i+1} + \frac{1}{2}(\varepsilon_{i+1}-\varepsilon_i)n_i n_{i+1}\right]~f(\varepsilon_i+\varepsilon_{i+1})}{\Omega_L(\bm{\nu})}\right\ra_s \, , \label{eq:curre2}
\een
\end{widetext}
after performing explicitly the averages with respect to $k_s$ and $\alpha_s$ (note in particular that $\la \alpha_s\ra=1/2=1- \la \alpha_s\ra$). We will be interested below in the large system size limit $L\to\infty$ where we expect the previous averages to become smooth functions of the diffusively-scaled space and time variables, $x$ and $t$, namely
\be
x(i)\equiv \frac{i}{L} \, , \qquad \Delta x \equiv x(i+1)-x(i) = \frac{1}{L} \, ,
\label{eq:space}
\ee
with $x\in[0,1]$ in the continuum limit, and 
\be
t(s)\equiv \frac{\la \tau_s\ra_s}{L^2} = L^{-3} \sum_{n=0}^{s-1} \la \delta \tau\ra_n \, ,
\label{eq:time}
\ee
where $\tau_s$ is the microscopic time at step $s$, and 
\be
\la \delta \tau\ra_n \equiv \lim_{L\to\infty} \la \Omega_L^{-1} \ra_n \, .
\label{eq:dt}
\ee
This average has a well-defined, finite value in the $L\to\infty$ limit, and defines a sort of microscopic time scale which depends explicitly on the choice of the collision rate function $f(\varepsilon_i+\varepsilon_{i+1})$ in Eq. (\ref{eq:CollisionRate}). In this diffusive scaling limit we hence expect the local average particle and energy densities, $\la n_{i}\ra_s$ and $\la \varepsilon_{i}\ra_s$, to be replaced by continuous fields $\rho(x,t)$ and $\varepsilon(x,t)\equiv \rho(x,t) T(x,t)$, respectively, with $T(x,t)$ a local temperature field, while the average particle and energy currents are expected to scale as
\be
\la j_{i}\ra_s \to \frac{\la \delta \tau\ra_s}{L^{2}} j(x,t) \, , \qquad \la q_{i}\ra_s \to \frac{\la \delta \tau\ra_s}{L^{2}} q(x,t) \, .
\label{eq:currfields}
\ee
This scaling, that will be demonstrated below, can be easily read from Eqs. (\ref{eq:currn2})-(\ref{eq:curre2}) by noting the \emph{discrete spatial derivatives} of particle and energy densities which appear in their numerator, that will give rise to an extra $L^{-1}$ scaling in the continuum limit, leading to an overall $L^{-2}$ scaling for the current fields. In this way, taking into account that $\la n_{i}\ra_{s+1} - \la n_{i}\ra_s \to \la \delta \tau\ra_s L^{-3} \partial_t \rho(x,t)$ and $\la j_{i-1} \ra_s - \la j_{i}\ra_s \to -\la \delta \tau\ra_s L^{-3} \partial_x j(x,t)$, and similarly for the energy balance, we arrive at two macroscopic balance equations for the particle and energy density (or equivalently temperature) fields in terms of their current fields
\ben
\partial_t \rho(x,t) + \partial_x j(x,t) &=& 0 \, , \label{eq:ncont} \\
\partial_t \left[\rho(x,t) T(x,t)\right] + \partial_x q(x,t) &=& 0 \, . \label{eq:econt} 
\een
The task remains to deduce the constitutive relations for the particle and energy current fields in terms of the density fields $\rho(x,t)$ and $T(x,t)$, starting from their microscopic expressions in Eqs. (\ref{eq:currn2})-(\ref{eq:curre2}). This challenge can be achieved within a local equilibrium approximation for the microscopic probability measure.

\subsection{Local equilibrium approximation}
\label{s3b}

Following Bogoliubov's picture on the emergence of hydrodynamic behavior in fluids \cite{zarate06a,de-groot13a}, we expect in the large system size limit $L\to\infty$ a strong separation of time scales between (a) a microscopic scale, of the order of a few typical times $\la \delta \tau\ra_s$, in which the system of interest relaxes \emph{locally} to an equilibrium-like state characterized by a local and instantaneous average particle density $\la n_{i}\ra_s$ and energy density $\la \varepsilon_{i}\ra_s$, and (b) a much longer macroscopic time scale over which these local average fields relax to their stationary values as dictated by the hydrodynamic equations (\ref{eq:ncont})-(\ref{eq:econt}). If this is the case, we expect that for times well beyond the microscopic time scale the probability measure $P(\bm{\nu};s)$ of a configuration $\bm{\nu}$ at step $s$ can be approximated by a local equilibrium probability measure with respect to the instantaneous local fields, the corrections to this approximation being weak, i.e. of order at most $L^{-1}$, namely
\be
P(\bm{\nu};s) \approx P_\text{LE}(\bm{\nu};s) + {\cal O}(L^{-1}) \, .
\label{eq:LE}
\ee
In what follows we will adopt this local equilibrium approximation to perform the averages in Eqs. (\ref{eq:currn2})-(\ref{eq:curre2}), neglecting the subdominant corrections to the leading hydrodynamic behavior. Note however that these weak ${\cal O}(L^{-1})$ corrections to local equilibrium are typically nonlocal and long-ranged \cite{derrida01a,derrida02a,derrida03a,bertini05b,hurtado10a}, being crucial to understand some of the key properties of nonequilibrium steady states \cite{bertini15a}. In this sense the previous approximation will be accurate as far as \emph{local} observables are concerned (as is the case here), but might fail when considering \emph{global} observables involving regions of size ${\cal O}(L)$ of the whole system \cite{derrida01a,derrida02a,derrida03a,bertini05b,hurtado10a}.

Let $\bm{\nu}_j\equiv (n_j,\varepsilon_j)$ be the local state of site $j$ when the total system is in configuration $\bm{\nu}$. The local equilibrium probability $P_\text{LE}(\bm{\nu};s)$ is a product measure
\be
P_\text{LE}(\bm{\nu};s) = \prod_{j=1}^L P_\text{LE}^{(j)}(\bm{\nu}_j;s) \, ,
\label{eq:LE2}
\ee
with $P_\text{LE}^{(j)}(\bm{\nu}_j;s)$ depending on $j$ and $s$ via the instantaneous local fields $\la n_j\ra_s$ and $\la T_j\ra_s\equiv \la\varepsilon_j\ra_s/\la n_j\ra_s$.
We may use Bayes theorem now to write
\be
P_\text{LE}^{(j)}(\bm{\nu}_j;s)\equiv P_\text{LE}^{(j)}(n_j,\varepsilon_j;s) = P_\text{LE}^{(j)}(n_j;s)~P_\text{LE}^{(j)}(\varepsilon_j|n_j;s) \, .
\label{eq:Bayes}
\ee
The local occupation number distribution can be simply written as
\be
P_\text{LE}^{(j)}(n_j;s)=\left\{ \begin{array}{ll}
\la n_j\ra_s  & \text{for }n_j=1\\
1-\la n_j\ra_s  & \text{for }\,n_j=0
\end{array}\right.
\ee
while the conditional energy probability distribution is
\be
P_\text{LE}^{(j)}(\varepsilon_j|n_j;s)=\left\{ \begin{array}{ll}
\la T_j\ra_s^{-1}~\text{e}^{-\varepsilon_j/\la T_j\ra_s} & \text{for } n_j=1\\
\delta(\varepsilon_j) & \text{for } n_j=0
\end{array}\right.
\ee
i.e. a local Gibbs measure with temperature $\la T_j\ra_s$ if site $j$ is occupied, or a Dirac delta-function at zero energy otherwise. The binary character of the occupation number variable can be now used to write $P_\text{LE}^{(j)}(\bm{\nu}_j;s)$ in a compact form
\begin{equation}
P_\text{LE}^{(j)}(\bm{\nu}_j;s)=n_j\frac{\la n_j\ra_s}{\la T_j\ra_s}~\text{e}^{-\varepsilon_j/\la T_j\ra_s}+(1-n_j)(1-\la n_j\ra_s)\delta(\varepsilon_j) .
\label{eq:LE3}
\end{equation}
Note that this local equilibrium approximation is a particular form of mean--field approximation where the possible spatial correlations between adjacent sites are neglected. As discussed above, these correlations are expected to be ${\cal O}(L^{-1})$ and hence can be safely neglected for local averages (though this point can be only settled in detailed numerical simulations testing the emerging hydrodynamic picture \cite{gutierrez-ariza19a}). In the next section we will use the local equilibrium picture here introduced to derive closed expressions for the average particle and energy currents, Eqs. (\ref{eq:currn2})-(\ref{eq:curre2}), in terms of the density and energy fields and their gradients.

\subsection{Constitutive relations and hydrodynamics}
\label{s3c}

In the course of this paper we will confront different averages with the common form
\be
I\equiv \frac{1}{L}\left\la \frac{g(\bm{\nu}_i,\bm{\nu}_{i+1})~f(\varepsilon_i+\varepsilon_{i+1})}{\Omega_L(\bm{\nu})} \right\ra_s \, ,
\label{eq:avegen}
\ee
with $g(\bm{\nu}_i,\bm{\nu}_{i+1})=g[(n_i,\varepsilon_i),(n_{i+1},\varepsilon_{i+1})]$ some function of the local state variables at pair $(i,i+1)$, see e.g. Eqs. (\ref{eq:currn2})-(\ref{eq:curre2}) for the average currents above. Let $\bm{\nu}_{\widehat{i,i+1}}$ be the microscopic configuration of a system with $L-2$ sites which results from configuration $\bm{\nu}$ after removing sites $i$ and $i+1$. The previous average can be written as
\begin{widetext}
\be
I = \frac{1}{L}\sum_{\bm{\nu}_i,\bm{\nu}_{i+1}} g(\bm{\nu}_i,\bm{\nu}_{i+1}) f(\varepsilon_i+\varepsilon_{i+1}) P_\text{LE}^{(i)}(\bm{\nu}_i;s) P_\text{LE}^{(i+1)}(\bm{\nu}_{i+1};s) ~\sum_{\bm{\nu}_{\widehat{i,i+1}}} \Omega_L^{-1}(\bm{\nu}) P_\text{LE}(\bm{\nu}_{\widehat{i,i+1}};s) \, ,
\label{eq:avegen2}
\ee
where we have already used the product form of the local equilibrium measure. The normalization factor $\Omega_L(\bm{\nu})$, defined in Eq. (\ref{eq:CollisionRate2}) and proportional to the total exit rate from configuration $\bm{\nu}$, can be now related with the normalization factor $\Omega_{L-2}(\bm{\nu}_{\widehat{i,i+1}})$ of the remnant configuration which results from $\bm{\nu}$ after removing sites $i$ and $i+1$. The latter can we written, see Eq. (\ref{eq:CollisionRate2}), as
\be
\Omega_{L-2}(\bm{\nu}_{\widehat{i,i+1}}) = \frac{1}{L-2} \left[ \sum_{\ell=1}^{i-2} \Theta(n_\ell+n_{\ell+1}) f\left(\varepsilon_\ell+\varepsilon_{\ell+1}\right) + \Theta(n_{i-1}+n_{i+2}) f\left(\varepsilon_{i-1}+\varepsilon_{i+2}\right) + \sum_{\ell=i+2}^{L} \Theta(n_\ell+n_{\ell+1}) f\left(\varepsilon_\ell+\varepsilon_{\ell+1}\right) \right]\, ,
\ee
where the possibility of a jump/collision event involving sites $i-1$ and $i+2$, which are now neighbors in the remnant configuration $\bm{\nu}_{\widehat{i,i+1}}$, is reflected in the middle of the previous equation. In this way, we find that
\be
\Omega_L(\bm{\nu}) = \frac{L-2}{L} \Omega_{L-2}(\bm{\nu}_{\widehat{i,i+1}}) + \frac{1}{L} \left[ \sum_{\ell=i-1}^{i+1} \Theta(n_\ell+n_{\ell+1}) f\left(\varepsilon_\ell+\varepsilon_{\ell+1}\right) -  \Theta(n_{i-1}+n_{i+2}) f\left(\varepsilon_{i-1}+\varepsilon_{i+2}\right) \right] \, ,
\ee
or equivalently $\Omega_L(\bm{\nu}) \approx \Omega_{L-2}(\bm{\nu}_{\widehat{i,i+1}}) + {\cal O}(L^{-1})$. Therefore, going back to the partial average appearing at the end of Eq. (\ref{eq:avegen2}), we have
\be
\sum_{\bm{\nu}_{\widehat{i,i+1}}} \Omega_L^{-1}(\bm{\nu}) P_\text{LE}(\bm{\nu}_{\widehat{i,i+1}};s) \approx  \la \Omega_{L-2}^{-1} \ra_s + {\cal O}(L^{-1}) \, ,
\label{eq:omega3}
\ee
which in the large size limit yields 
\be
\lim_{L\to\infty} \la \Omega_{L-2}^{-1} \ra_s = \lim_{L\to\infty} \la \Omega_{L}^{-1} \ra_s = \la \delta\tau\ra_s \, , 
\label{eq:omega4}
\ee
i.e. the microscopic time scale defined in Eq. (\ref{eq:dt}). In this limit, the average of interest (\ref{eq:avegen}) hence boils down to the following two-body problem
\be
I = \frac{\la \delta\tau\ra_s}{L} \left\la g(\bm{\nu}_i,\bm{\nu}_{i+1}) f(\varepsilon_i+\varepsilon_{i+1}) \right\ra_s \, ,
\label{eq:avegen3}
\ee
with
\ben
\left\la g(\bm{\nu}_i,\bm{\nu}_{i+1}) f(\varepsilon_i+\varepsilon_{i+1}) \right\ra_s &=& \sum_{\bm{\nu}_i,\bm{\nu}_{i+1}} ~g(\bm{\nu}_i,\bm{\nu}_{i+1}) ~f(\varepsilon_i+\varepsilon_{i+1}) P_\text{LE}^{(i)}(\bm{\nu}_i;s) ~P_\text{LE}^{(i+1)}(\bm{\nu}_{i+1};s)  \\
&=& \sum_{n_i,n_{i+1}=0,1} \int_0^\infty d\varepsilon_i d\varepsilon_{i+1}  ~g[(n_i,\varepsilon_i),(n_{i+1},\varepsilon_{i+1})] ~f(\varepsilon_i+\varepsilon_{i+1})~P_\text{LE}^{(i)}(n_i,\varepsilon_i;s) P_\text{LE}^{(i+1)}(n_{i+1},\varepsilon_{i+1};s) \, . \nonumber
\label{eq:avegen4}
\een
where we have made explicit the dependence on the state variables in the second line of the equation.

To further proceed, we now explicitly apply the local equilibrium approximation of the previous section to compute this average, that will depend on the local density and temperature fields evaluated at the sites of interest, $(i,i+1)$, see Eq. (\ref{eq:LE3}). In order to be consistent with the continuum limit introduced in Section \S\ref{s3a}, we now assume that the local changes in the density and temperature fields across infinitesimally separated regions are small, namely $\la n_{i+1}\ra_s - \la n_{i}\ra_s ={\cal O}(L^{-1})$ and $\la T_{i+1}\ra_s - \la T_{i}\ra_s ={\cal O}(L^{-1})$. In this way, recalling that the spatial separation between nearby points in the diffusive scale is $\Delta x=1/L$, see Eq. (\ref{eq:space}), we can write
\be
\la n_{i+1}\ra_s = \la n_{i}\ra_s +  \frac{1}{L} \frac{\la n_{i+1}\ra_s - \la n_{i}\ra_s}{\Delta x} \underset{L\gg 1}{\approx} \rho + \frac{1}{L} \partial_x\rho \, , \label{eq:rhodiff}  
\ee
where $\rho=\rho(x,t)$ with $x=i/L$ and $t=\la\tau_s\ra_s/L^2$, and similarly $\la T_{i+1}\ra_s\approx T + \frac{1}{L} \partial_x T$. We now may use these expressions to expand the local equilibrium measure at site $i+1$ up to first order in $L^{-1}$, see Eqs. (\ref{eq:LE3}) and (\ref{eq:avegen4}), 
\be
P_\text{LE}^{(i+1)}(n_{i+1},\varepsilon_{i+1};s)=n_{i+1} \text{e}^{-\varepsilon_{i+1}/T}\frac{\rho}{T}\left[1+\frac{\partial_x\rho}{L\rho} -\left(1-\frac{\varepsilon_{i+1}}{T} \right)\frac{\partial_x T}{L T}\right] + (1-n_{i+1})(1-\rho)\delta(\varepsilon_{i+1}) .
\label{eq:LE4}
\ee
We hence find that the local equilibrium probability of a given state for the pair $(i,i+1)$ can be naturally splitted as
\ben
P_\text{LE}^{(i)}(\bm{\nu}_i;s) ~P_\text{LE}^{(i+1)}(\bm{\nu}_{i+1};s) &=& n_{i}n_{i+1}P_{LE}^{\tiny \newmoon\newmoon}(\varepsilon_{i},\varepsilon_{i+1};s)+n_{i}(1-n_{i+1})P_{LE}^{\tiny\newmoon\fullmoon}(\varepsilon_{i},\varepsilon_{i+1};s) \nonumber \\
&+& (1-n_{i})n_{i+1}P_{LE}^{\tiny\fullmoon\newmoon}(\varepsilon_{i},\varepsilon_{i+1};s)+(1-n_{i})(1-n_{i+1})P_{LE}^{\tiny\fullmoon\fullmoon}(\varepsilon_{i},\varepsilon_{i+1};s) \, ,
 \label{eq:pLE0}
\een
with the definitions
\ben
P_{LE}^{\tiny \newmoon\newmoon}(\varepsilon_{i},\varepsilon_{i+1};s)&=&\frac{\rho^{2}}{T^{2}} \left[1+\frac{\partial_{x}\rho}{L\rho}-\left(1-\frac{\varepsilon_{i+1}}{T}\right)\frac{\partial_{x}T}{TL} \right] \text{e}^{-(\varepsilon_{i}+\varepsilon_{i+1})/T} \label{eq:pLE1} \\
P_{LE}^{\tiny\newmoon\fullmoon}(\varepsilon_{i},\varepsilon_{i+1};s)&=&\frac{\rho}{T}  \left(1-\rho-\frac{\partial_{x}\rho}{L}\right) \text{e}^{-\varepsilon_{i}/T} \delta\left(\varepsilon_{i+1}\right) \label{eq:pLE2} \\
P_{LE}^{\tiny\fullmoon\newmoon}(\varepsilon_{i},\varepsilon_{i+1};s)&=&\frac{\rho}{T}\left(1-\rho\right)\left[ 1+\frac{\partial_{x}\rho}{L\rho}-\left(1-\frac{\varepsilon_{i+1}}{T}\right)\frac{\partial_{x}T}{TL}\right] \delta\left(\varepsilon_{i}\right) \text{e}^{-\varepsilon_{i+1}/T} \label{eq:pLE3} \\
P_{LE}^{\tiny\fullmoon\fullmoon}(\varepsilon_{i},\varepsilon_{i+1};s)&=& \left(1-\rho\right)\left(1-\rho-\frac{\displaystyle \partial_x\rho}{\displaystyle L}\right)\delta\left(\varepsilon_{i}\right)\delta\left(\varepsilon_{i+1}\right) \label{eq:pLE4} 
\label{eq:probabilidades_de_EL}
\een
\end{widetext}
where we recall again that $\rho=\rho(x,t)$ and $T=T(x,t)$. We are now in position to compute explicitly the average particle and energy currents within the local equilibrium approximation, see Eqs. (\ref{eq:currn2})-(\ref{eq:curre2}). As a consequence of the previous splitting of the local equilibrium probability measure, we can write the average particle current, see Eqs. (\ref{eq:currn2}) and (\ref{eq:avegen4}), as
\be
\la j_{i}\ra_s = \la j_{i}\ra_s^{\tiny\newmoon\fullmoon} + \la j_{i}\ra_s^{\tiny\fullmoon\newmoon} \, ,
\label{eq:currn3}
\ee
since $ \la j_{i}\ra_s^{\tiny \newmoon\newmoon} = 0 =  \la j_{i}\ra_s^{\tiny\fullmoon\fullmoon}$ due to the $(n_i-n_{i+1})$ term in Eq. (\ref{eq:currn2}). The first of the two non-zero contributions to the average particle current is
\ben
\la j_{i}\ra_s^{\tiny\newmoon\fullmoon} &=& \frac{\la \delta\tau\ra_s}{L} \int_0^\infty d\varepsilon_i d\varepsilon_{i+1} f(\varepsilon_i +\varepsilon_{i+1})  P_{LE}^{\tiny\newmoon\fullmoon}(\varepsilon_{i},\varepsilon_{i+1};s) \nonumber \\
&=& \frac{\la \delta\tau\ra_s}{L} \rho \left(1-\rho-\frac{\partial_{x}\rho}{L}\right) F_0(T) \, ,
\label{eq:j10}
\een
see Eq. (\ref{eq:pLE2}), where we have defined a generic integral
\be
F_n(T) \equiv \int_0^\infty dy~\text{e}^{-y} y^n f(Ty) \, ,
\label{eq:F1}
\ee
after a change of variables $y=\varepsilon_i/T$ in the last equality. Similarly
\ben
\la j_{i}\ra_s^{\tiny\fullmoon\newmoon} &=& -\frac{\la \delta\tau\ra_s}{L} \rho \left(1-\rho\right)\left[ \left(1+\frac{\partial_{x}\rho}{L\rho}-\frac{\partial_{x}T}{LT}\right)F_0(T) \right. \nonumber \\
&+& \left.  \frac{\partial_{x}T}{LT} F_1(T) \right] \, . 
\label{eq:j01} 
\een
Note that a simple integration by parts allows to relate the functions $F_n(T)$ and $F_{n+1}(T)$ in a recurrent manner, namely
\be
F_{n+1}(T) = (n+1) F_n(T) + T F'_n(T) \, ,
\label{eq:relF12}
\ee
with $'$ denoting derivative with respect to the argument. Putting all together, we find for the average particle current
\be
\la j_{i}\ra_s = \frac{\la \delta\tau\ra_s}{L^2} \left[-F_0(T) \partial_x\rho - \rho(1-\rho) F'_0(T)  \partial_x T \right] \, ,
\label{eq:currn4}
\ee
which confirms the heuristic scaling anticipated in Eq. (\ref{eq:currfields}). In this way, the constitutive relation for the particle current field in the diffusive scaling limit is just
\be
j(x,t) = -D_{11}(T) \partial_x \rho(x,t) - D_{12}(\rho,T) \partial_x T(x,t) \, ,
\label{eq:currn5}
\ee
with the following transport coefficients
\ben
D_{11}(T) &=& F_0(T) \, , \label{eq:D11} \\
D_{12}(\rho,T) &=& \rho(1-\rho) F'_0(T) \, , \label{eq:D12}
\een
which depend nonlinearly on the local particle density and temperature fields, $\rho(x,t)$ and $T(x,t)$ respectively, and on the function $f(\varepsilon)$ controlling the energy dependence of the microscopic jump/collision rate. Note that the constitutive relation (\ref{eq:currn5}) captures the possibility of a particle flow in the absence of a density gradient, due exclusively to a temperature gradient. This is the well-known Soret effect \cite{de-groot13a}.

An equivalent calculation for the average energy current (\ref{eq:curre2}), summarized in Appendix \ref{app1}, leads to $\la q_{i}\ra_s = \la \delta\tau\ra_s L^{-2} q(x,t)$, with
\be
q(x,t) = -D_{21}(T) \partial_x \rho(x,t) - D_{22}(\rho,T) \partial_x T(x,t) \, ,
\label{eq:curre5}
\ee
with the additional transport coefficients
\ben
D_{21}(T) &=& T F_1(T) \, , \label{eq:D21} \\
D_{22}(\rho,T) &=& \rho(1-\rho)[F_2(T)-F_1(T)] + \frac{\rho^2}{12} F_3(T) . \, \label{eq:D22}
\een
This shows that an energy current due exclusively to the presence of a density gradient (Dufour effect \cite{de-groot13a}) may appear in the kinetic exclusion process. The previous constitutive relations for the particle and energy current fields can be written in an unified way using vector notation, namely
\be
\left(\begin{array}{l}
j\\
q
\end{array}\right)=-\mathbb{D}(\rho,T) \left(\begin{array}{l}
\partial_{x}\rho\\
\partial_{x}T
\end{array}\right) \, , \label{eq:FickFourier}
\ee
with a transport coefficients matrix
\be
\mathbb{D}(\rho,T)=\left(\begin{array}{cc}
D_{11}(T) & D_{12}(\rho,T)\\
D_{21}(T) & D_{22}(\rho,T)
\end{array}\right)\, ,
\label{eq:Dmatrix}
\ee
This is just a generalized Fick-Fourier's law \cite{de-groot13a} which, together with the continuity equations (\ref{eq:ncont})-(\ref{eq:econt}) for the density and temperature fields, result in the following hydrodynamic equations for the kinetic exclusion process
\be
\partial_t \left(\begin{array}{c}
\rho\\
\rho T
\end{array}\right) - \mathbb{D}(\rho,T) \partial_x \left(\begin{array}{c}
\rho\\
T
\end{array}\right) = 0 \, .
\ee

\subsection{Entropy production and Onsager matrix}
\label{s3d}

In accordance with our local equilibrium hypothesis, we expect thermodynamic relations to hold \emph{locally} in our system. In particular, we expect a local version of the Gibbs relation linking variations of entropy $\tilde{s}$, internal energy $\varepsilon=\rho T$, specific volume $v$ and density $\rho$ to remain valid \cite{de-groot13a}, i.e.
\be
d \varepsilon =T d\tilde{s} - p dv + \mu d\rho \,
\label{eq:Gibbs}
\ee
with $\mu$ the chemical potential and $p$ the pressure. Taking into account that the size of our system is kept fixed at any moment, so $dv=0$, we can write
\be
\partial_t \tilde{s} = \frac{1}{T} \partial_t (\rho T) -\frac{\mu}{T} \partial_t \rho \, ,
\label{eq:Gibbs2}
\ee
and using here the continuity equations (\ref{eq:ncont})-(\ref{eq:econt}) derived in previous sections for the local energy and density fields, we find
\ben
\partial_t \tilde{s} &=& - \frac{1}{T} \partial_x q + \frac{\mu}{T} \partial_x j \nonumber \\ 
&=& -\partial_x\left(\frac{q}{T} -  \frac{\mu}{T} j\right) + q \partial_x \left( \frac{1}{T} \right) + j \partial_x \left(- \frac{\mu}{T} \right) \, ,
\label{eq:entropybalance}
\een
where we have used the chain rule in the second equality. This expression has the typical form of an entropy balance equation $\partial_t \tilde{s} = -\partial_x J_{\tilde{s}} + \sigma$, which allows us to identify the entropy flux $J_{\tilde{s}}$ from the environment and the entropy production $\sigma$ due to the irreversible phenomena occurring within the system \cite{de-groot13a},
\ben
J_{\tilde{s}} &=& \frac{q}{T} -  \frac{\mu}{T} j \label{eq:sflux} \\
\sigma &=& q \partial_x \left( \frac{1}{T} \right) + j \partial_x \left(- \frac{\mu}{T} \right) \, . \label{eq:sprod}
\een
Note that the entropy production term has the standard form $\sigma = \sum_k j_k X_k$, i.e. it is proportional to the currents of different magnitudes times their corresponding \emph{thermodynamic forces}, which can be now readily identified from Eq. (\ref{eq:sprod}). In particular, $X_1=\partial_x (-\mu/T)$ and $X_2=\partial_x (1/T)$. Moreover, the currents are in turn proportional to these thermodynamic forces, $j_k=\sum_\ell L_{k\ell} X_\ell$, so the entropy production becomes a quadratic form of the thermodynamic forces, $\sigma=\sum_{k,\ell} L_{k\ell} X_k X_\ell$. The factors $L_{k\ell}(\rho,T)$ define the well-known Onsager matrix $\mathbb{L}(\rho,T)$ of phenomenological coefficients \cite{de-groot13a}.

In order to identify the Onsager matrix, we need to write the particle and energy currents in terms of the thermodynamic forces $X_1=\partial_x (-\mu/T)$ and $X_2=\partial_x (1/T)$, and to do so we need to compute the chemical potential $\mu$ for the kinetic exclusion process. This can be achieved in a number of different ways, the simplest for our purposes here being $-\mu/T = (\partial \tilde{s}/\partial \rho)_{\varepsilon,v}$, see Eq. (\ref{eq:Gibbs}). The entropy density $\tilde{s}$ can be calculated from the Gibbs entropy expression $\tilde{s}=-\sum_{\bm{\nu}_i} P_\text{LE}^{(i)}(\bm{\nu}_i;s) \ln P_\text{LE}^{(i)}(\bm{\nu}_i;s)$, leading to
\be
\tilde{s} = -(1-\rho) \ln (1-\rho) - 2 \rho \ln \rho + \rho \ln \varepsilon + \rho \, .
\label{eq:entropy}
\ee
In this way
\be
-\frac{\mu}{T} = \ln \left[\frac{T(1-\rho)}{\rho} \right] \, ,
\label{eq:mu}
\ee
so the thermodynamic force associated to the particle density is $X_1= \partial_x (-\mu/T) = T^{-1} \partial_x T  - [\rho(1-\rho)]^{-1}\partial_x\rho$. We hence find
\ben
j &=& -\frac{L_{11}}{\rho(1-\rho)} \partial_x \rho - \left(\frac{L_{12}}{T^2} - \frac{L_{11}}{T}\right) \partial_x T \, , \nonumber \\
q &=& -\frac{L_{21}}{\rho(1-\rho)} \partial_x \rho - \left(\frac{L_{22}}{T^2} - \frac{L_{21}}{T}\right) \partial_x T \, , \nonumber
\een
and comparing these expressions with the Fick-Fourier's law derived in the previos section, see Eqs. (\ref{eq:currn5}) and (\ref{eq:curre5}), we arrive at
\ben
L_{11}(\rho,T) &=& \rho(1-\rho) F_0(T) \nonumber \\
L_{12}(\rho,T) &=& \rho(1-\rho) T F_1(T) = L_{21}(\rho,T)\label{eq:Onsager} \\
L_{22}(\rho,T) &=& \rho(1-\rho) T^2 F_2(T) + \frac{\rho^2 T^2}{12} F_3(T) \, , \nonumber
\een
which define the phenomenological coefficients of Onsager (symmetric) matrix. 

To ensure the positivity of the entropy production, $\sigma\ge 0$, as demanded by the second law, the coefficients of Onsager matrix must obey certain restrictions \cite{de-groot13a}. As the entropy production is a quadratic form, $\sigma=\sum_{k,\ell} L_{k\ell} X_k X_\ell$, sufficient conditions to guarantee its positivity \cite{de-groot13a} are $L_{ii}\ge 0$ for $i=1,2$ and $L_{11}L_{22} \ge \frac{1}{4}(L_{12}+L_{21})^2$, which lead to 
\ben
F_0(T) &\ge& 0 \, , \label{eq:cond1} \\
F_2(T) + \frac{\rho}{12(1-\rho)} F_3(T) &\ge& 0  \, , \label{eq:cond2} \\
F_0(T) \left[F_2(T) + \frac{\rho}{12(1-\rho)} F_3(T) \right] &\ge& F_1(T)^2  \, . \label{eq:cond3} 
\een
The first two conditions above, Eqs. \eqref{eq:cond1}-\eqref{eq:cond2}, are automatically verified once taken into account the definition (\ref{eq:F1}) of the functions $F_n(T)$, the positivity of the function $f(\varepsilon)$, and the density domain, $0\le \rho\le 1$. On the other hand, condition \eqref{eq:cond3} must be fulfilled $\forall~T,\rho$ to guarantee a positive entropy production term. In this way, noticing that $\rho/(1-\rho)$ is a monotonously increasing function of the density in the interval $0\le \rho\le 1$, Eq. \eqref{eq:cond3} will be satisfied $\forall~T,\rho$ if and only if
\be
F_0(T) F_2(T) \ge F_1(T)^2  \, \qquad \forall T ,
\label{eq:cond4}
\ee
or equivalently $F_2(T)/F_0(T)\ge [F_1(T)/F_0(T)]^2$. Noting that $F_n(T)/F_0(T)$ is nothing but the $n$-th moment of the normalized pdf $\exp(-y)\, f(Ty)/F_0(T)$, condition \eqref{eq:cond4} just expresses the possitivity of the variance of such pdf, and hence it is automatically satisfied for all functions $f(\varepsilon)$ leading to finite values for the integral $F_n(T)$, $n\le 2$. This proves the positivity of entropy production in the kinetic exclusion process.

\section{Hydrodynamic fluctuations}
\label{s4}

Up to now we have obtained the macroscopic hydrodynamic equations governing the evolution of the kinetic exclusion process for sufficiently large spatial and temporal scales. We now want to characterize the unavoidable local fluctuations of the current fields which accompany this evolution, i.e. we want to deduce from the microscopic dynamics and within the local equilibrium approximation the fluctuating hydrodynamics which describe the kinetic exclusion process at a \emph{mesoscopic} level. These fluctuations will appear as noise terms perturbing the local current fields, and we will argue below that such noises are white and Gaussian. Moreover, these noises will be shown to be weak, in the sense that their amplitudes scale as $\sim{\cal O}(L^{-1/2})$ in the large system size limit $L\gg 1$.

In order to proceed, we first have to split the microscopic currents $j_{i,s}$ and $q_{i,s}$ of Section \S\ref{s3a}, see Eqs. \eqref{eq:currn} and \eqref{eq:curre}, into some main terms, $\bar{j}_{i,s}$ and $\bar{q}_{i,s}$ respectively, and their corresponding noises, $\xi_{i,s}$ and $\zeta_{i,s}$, i.e.
\ben
j_{i,s} &=& \bar{j}_{i,s} + \xi_{i,s} \, , \label{eq:xi} \\
q_{i,s} &=& \bar{q}_{i,s} + \zeta_{i,s} \label{eq:zeta}\, . 
\een
The main terms in the previous splitting must be \emph{configurational observables}, i.e. sole functions of the local state variables (occupation numbers and energies), and independent of the jump/collision parameters $(k_s,\alpha_s)$ at step $s$. Moreover, their average over trajectories must coincide with that of the microscopic currents, i.e. $\la j_{i} \ra_s=\la \bar{j}_{i}\ra_s$ and $\la q_{i} \ra_s=\la \bar{q}_{i}\ra_s$. It is clear from Eqs. \eqref{eq:currn2} and \eqref{eq:curre2} that the choice
\ben
\bar{j}_{i,s} &\equiv& \frac{1}{L\Omega_L(\bm{\nu})} (n_{i,s}-n_{i+1,s})~f(\varepsilon_{i,s}+\varepsilon_{i+1,s}) \, , \label{eq:currnmic} \\
\bar{q}_{i,s} &\equiv& \frac{1}{L\Omega_L(\bm{\nu})} \Big[n_{i,s}\varepsilon_{i,s} -n_{i+1,s}\varepsilon_{i+1,s}  \label{eq:curremic} \\
&+&  \frac{1}{2}(\varepsilon_{i+1,s}-\varepsilon_{i,s})n_{i,s} n_{i+1,s}\Big]~f(\varepsilon_{i,s}+\varepsilon_{i+1,s}) \, , \nonumber
\een
guarantees these constraints on the averages. This is nothing but the microscopic version of Fick-Fourier's law \eqref{eq:FickFourier} expressing the proportionality between the microscopic particle and heat currents and the associated \emph{instantaneous} density and energy local gradients. It is important to stress the differences between the exact microscopic currents $j_{i,s}$ and $q_{i,s}$ and the main terms $\bar{j}_{i,s}$ and $\bar{q}_{i,s}$ in the splitting of Eqs. \eqref{eq:xi}-\eqref{eq:zeta}. Indeed, while $j_{i,s}$ and $q_{i,s}$ are exactly zero unless a jump/collision event happens at pair $(i,i+1)$ at time step $s$, the values of $\bar{j}_{i,s}$ and $\bar{q}_{i,s}$ may take a non-trivial, non-zero value even if nothing happens at pair $(i,i+1)$ at time step $s$. 

We want to study the statistical properties of the noise terms $\xi_{i,s}=j_{i,s}-\bar{j}_{i,s}$ and $\zeta_{i,s}=q_{i,s}-\bar{q}_{i,s}$ as captured by e.g. their average value and correlation matrix. From the constraints on the averages, $\la j_{i} \ra_s=\la \bar{j}_{i}\ra_s$ and $\la q_{i} \ra_s=\la \bar{q}_{i}\ra_s$, it is clear that
\be
\la \xi_i\ra_s = 0 \, , \qquad \la \zeta_i\ra_s = 0 \, ,
\label{eq:noiseav}
\ee
so the noises do not contribute to the average currents, as expected. On the other hand, the two-body correlators for the noises can be simply written as
\ben
\la \xi_{i,s} \xi_{\ell,s'}\ra &=& \la j_{i,s} j_{\ell,s'} \ra - \la \bar{j}_{i,s} \bar{j}_{\ell,s'}\ra \, , \label{eq:noisecorr1} \\
\la \zeta_{i,s} \zeta_{\ell,s'}\ra &=& \la q_{i,s} q_{\ell,s'} \ra - \la \bar{q}_{i,s} \bar{q}_{\ell,s'}\ra \, , \label{eq:noisecorr2} \\
\la \xi_{i,s} \zeta_{\ell,s'}\ra &=& \la j_{i,s} q_{\ell,s'} \ra - \la \bar{j}_{i,s} \bar{q}_{\ell,s'}\ra \, . \label{eq:noisecorr3} 
\een
Using now the microscopic definition of the particle and energy currents, $j_{i,s}$ and $q_{i,s}$ in Eqs.  \eqref{eq:currn} and \eqref{eq:curre}, one can easily prove that the above correlators vanish when evaluated at different time steps, so that $\la \xi_{i,s} \xi_{\ell,s'}\ra = \la \zeta_{i,s} \zeta_{\ell,s'}\ra = \la \xi_{i,s} \zeta_{\ell,s'}\ra = 0$ $\forall s\neq s'$, meaning that the particle and energy current noises at different times are uncorrelated. We hence fix from now on $s'=s$ unless otherwise specified. In this particular case, a glance at the definitions of the main contributions to the currents, Eqs. \eqref{eq:currnmic}-\eqref{eq:curremic}, shows that the second terms in the rhs of the Eqs. \eqref{eq:noisecorr1}-\eqref{eq:noisecorr3} are of order $\la \bar{j}_{i} \bar{j}_{\ell}\ra_s\sim \la \bar{q}_{i} \bar{q}_{\ell}\ra_s\sim\la \bar{j}_{i} \bar{q}_{\ell}\ra_s\sim{\cal O}(L^{-2})$. We will show below that, in contrast, the first terms in the rhs of these equations scale as $\sim{\cal O}(L^{-1})$, rendering negligible the former against the latter in the $L\gg 1$ limit. Therefore the calculation of the noise correlators boils dow to computing the averages $\la {j}_{i} {j}_{\ell}\ra_s$, $\la {q}_{i} {q}_{\ell}\ra_s$, and $\la {j}_{i} {q}_{\ell}\ra_s$.

We start with the simplest case, i.e. that of the particle current noise correlator $\la \xi_{i} \xi_{\ell}\ra_s= \la {j}_{i} {j}_{\ell}\ra_s + {\cal O}(L^{-2})$. Using now the microscopic expression for ${j}_{i,s}$, see Eq. \eqref{eq:currn}, we have that
\ben
{j}_{i,s} {j}_{\ell,s} &=& \delta_{k_s,i} \delta_{k_s,\ell} \left(n_{i,s} - n_{i+1,s} \right) \left(n_{\ell,s} - n_{\ell+1,s} \right) \nonumber \\
&=& \delta_{i,\ell} \delta_{k_s,i} \left(n_{i,s} - n_{i+1,s} \right)^2 \, ,
\label{eq:correlpp1}
\een
where we recall that index $k_s$ indicates the (random) pair where a jump/collision event happens at time step $s$. Note that in the second identity we have used the fact that this product of particle currents is exactly zero unless $i=\ell$ (otherwise at least one of the two microscopic particle currents will be exactly zero). As the occupation number variables can take only two discrete values, $n_i=0,1$, the microscopic squared particle current $\left(n_{i,s} - n_{i+1,s} \right)^2=0,1$, so we can substitute $\left(n_{i,s} - n_{i+1,s} \right)^2 = \left \vert n_{i,s} - n_{i+1,s}\right \vert$. Averaging now over all possible trajectories, see Eq. \eqref{eq:ave}, we thus find
\be
\la j_{i} j_{\ell} \ra_s = \delta_{i,\ell} \la |j_{i}|\ra_s = \delta_{i,\ell} \left[\la j_{i}\ra_s^{\tiny\newmoon\fullmoon} - \la j_{i}\ra_s^{\tiny\fullmoon\newmoon} \right] \, ,
\label{eq:correlpp2}
\ee
where we have used in the last equality the decomposition of the particle current introduced in Section \S\ref{s3c} as a consequence of the splitting of the local equilibrium measure $P_\text{LE}^{(i)}(\bm{\nu}_i;s) P_\text{LE}^{(i+1)}(\bm{\nu}_{i+1};s)$ in terms of the possible local pair configurations, see Eqs. \eqref{eq:pLE1}-\eqref{eq:pLE4}. Notice in particular that the (negative) current $\la j_{i}\ra_s^{\tiny\fullmoon\newmoon}$ associated to a particle jump to the left enters with a minus sign in the above expression as a consequence of the above absolute value, $\la j_{i} j_{\ell} \ra_s = \delta_{i,\ell} \la |j_{i}|\ra_s$. This detail is responsible of defining the leading value of this correlator. In fact, we may use now the expressions already derived for $\la j_{i}\ra_s^{\tiny\newmoon\fullmoon}$ and $\la j_{i}\ra_s^{\tiny\fullmoon\newmoon}$ in Section \S\ref{s3c}, see Eqs \eqref{eq:j10} and \eqref{eq:j01}, to arrive at the following formula for the particle current noise correlator in terms of the local density and temperature fields,
\be
\la \xi_{i} \xi_{\ell} \ra_s = \delta_{i,\ell} \frac{\la \delta\tau\ra_s}{L} 2 \rho (1-\rho) F_0(T) + {\cal O}(L^{-2}) \, .
\label{eq:correlpp3}
\ee
Here we have already neglected ${\cal O}(L^{-2})$ gradient corrections against the dominant ${\cal O}(L^{-1})$ terms. Notice that the previous expression confirms the ${\cal O}(L^{-1})$ scaling of the correlators $\la j_{i} j_{\ell} \ra_s$ anticipated above, validating \emph{a posteriori} our analysis.

Next we focus on the cross-correlation between the particle and energy current noises, $\la \xi_{i,s} \zeta_{\ell,s}\ra = \la j_{i,s} q_{\ell,s} \ra + {\cal O}(L^{-2})$. In this case, using the definitions of the microscopic currents Eqs. \eqref{eq:currn} and \eqref{eq:curre}, and reasoning along the same lines, we have
\ben
{j}_{i,s} {q}_{\ell,s} &=& \delta_{i,\ell}\delta_{k_s,i} \left(n_{i,s} - n_{i+1,s} \right)  \left[n_{i,s} \varepsilon_{i,s} - n_{i+1,s} \varepsilon_{i+1,s} \right. \nonumber \\
&& + \left. \left((1-\alpha_s) \varepsilon_{i+1,s} - \alpha_s \varepsilon_{i,s} \right)n_{i,s} n_{i+1,s}\right] \nonumber \\
&=& \delta_{i,\ell}\left(n_{i,s}\varepsilon_{i,s} +  n_{i+1,s} \varepsilon_{i+1,s} \right) \, ,
\label{eq:correlpe1}
\een
where, as above, we have made explicit that this product of particle and energy currents is exactly zero unless $i=\ell$, as well as the fact that the particle-particle collision contribution to this cross-product always vanish since $n_{i,s} n_{i+1,s}=1$ necessarily implies that $(n_{i,s} - n_{i+1,s} )=0$. Taking now averages over trajectories, and recalling the decomposition of the energy current introduced in Section \S\ref{s3c} resulting from the splitting of the local equilibrium measure, we find
\ben
\la \xi_{i,s} \zeta_{\ell,s}\ra &=& \delta_{i,\ell}\left[\la q_{i}\ra_s^{\tiny\newmoon\fullmoon} - \la q_{i}\ra_s^{\tiny\fullmoon\newmoon} \right] \label{eq:correlpe2} \\
&=& \delta_{i,\ell} \frac{\la \delta\tau\ra_s}{L} 2 \rho (1-\rho) T F_1(T) + {\cal O}(L^{-2}) \, , \nonumber
\een
where again we have neglected ${\cal O}(L^{-2})$ gradient corrections against the dominant ${\cal O}(L^{-1})$ terms. This expression confirms once more the scaling anticipated above.

We proceed now with the energy current noise correlator $\la \zeta_{i,s} \zeta_{\ell,s}\ra = \la q_{i,s} q_{\ell,s} \ra + {\cal O}(L^{-2})$. From the microscopic expression \eqref{eq:curre}
\ben
q_{i,s} q_{\ell,s} &=& \delta_{i,\ell}\delta_{k_s,i} \left[n_{i,s} \varepsilon_{i,s} - n_{i+1,s} \varepsilon_{i+1,s} \right. \label{eq:correlee1} \\
&+& \left. \left((1-\alpha_s) \varepsilon_{i+1,s} - \alpha_s \varepsilon_{i,s} \right)n_{i,s} n_{i+1,s} \right]^2 \, . \nonumber
\een
Averaging now over trajectories and using the same splitting as above, we obtain
\be
\la q_{i,s} q_{\ell,s} \ra = \delta_{i,\ell} \left[ \la q_{i}^2\ra_s^{\tiny\newmoon\fullmoon} + \la q_{i}^2\ra_s^{\tiny\fullmoon\newmoon} + \la q_{i}^2\ra_s^{\tiny\newmoon\newmoon}\right] \, ,
\label{eq:correlee2}
\ee
and these averages can be calculated now as particular cases of the general two-body integral \eqref{eq:avegen4} introduced in Section \S\ref{s3c}. For the first term we find
\ben
\la q_{i}^2\ra_s^{\tiny\newmoon\fullmoon} &=& \frac{\la \delta\tau\ra_s}{L} \int_0^\infty d\varepsilon_i d\varepsilon_{i+1}~\varepsilon_i^2 f(\varepsilon_i +\varepsilon_{i+1})  P_{LE}^{\tiny\newmoon\fullmoon}(\varepsilon_{i},\varepsilon_{i+1};s) \nonumber \\
&=& \frac{\la \delta\tau\ra_s}{L} \rho(1-\rho) T^2 F_2(T) + {\cal O}(L^{-2}) \, ,
\label{eq:correlee3a}
\een
where we have used Eq. \eqref{eq:pLE2} in the second equality, disregarding already subdominant ${\cal O}(L^{-2})$ terms. An equivalent calculation for the second term in Eq. \eqref{eq:correlee2} leads to the same result, namely
\be
\la q_{i}^2\ra_s^{\tiny\fullmoon\newmoon} = \frac{\la \delta\tau\ra_s}{L} \rho(1-\rho) T^2 F_2(T) + {\cal O}(L^{-2}) = \la q_{i}^2\ra_s^{\tiny\newmoon\fullmoon} \, .
\label{eq:correlee3b}
\ee
The contribution of particle collisions to the energy current correlator is on the other hand somewhat more involved. In this case, by taking $n_{i,s}=1=n_{i+1,s}$ in the general expression for $q_{i,s}$, we have that
\begin{widetext}
\ben
\la q_{i}^2\ra_s^{\tiny\newmoon\newmoon} &=& \frac{\la \delta\tau\ra_s}{L} \int_0^\infty d\varepsilon_i d\varepsilon_{i+1} \int_0^1 d\alpha_s \left[(1-\alpha_s) \varepsilon_{i,s} - \alpha_s \varepsilon_{i+1,s} \right]^2 f(\varepsilon_i +\varepsilon_{i+1})  P_{LE}^{\tiny\newmoon\newmoon}(\varepsilon_{i},\varepsilon_{i+1};s) \nonumber \\
&=& \frac{\la \delta\tau\ra_s}{L} \frac{1}{3}\frac{\rho^2}{T^2} \int_0^\infty d\varepsilon_i d\varepsilon_{i+1} (\varepsilon_i^2 +\varepsilon_{i+1}^2 - \varepsilon_i \varepsilon_{i+1}) f(\varepsilon_i +\varepsilon_{i+1}) \text{e}^{-(\varepsilon_i +\varepsilon_{i+1})/T} + {\cal O}(L^{-2}) \, ,
\label{eq:correlee3c0}
\een
\end{widetext}
where we have used Eq. \eqref{eq:pLE1}, neglecting as before ${\cal O}(L^{-2})$ terms, together with the averages $\la \alpha_s^2\ra = 1/3 = \la (1-\alpha_s)^2\ra$ and $\la \alpha_s(1-\alpha_s)\ra = 1/6$. The last integral can be solved in simple terms using a change of variables to polar coordinates, see Appendix \ref{app2}, and yields a value $\frac{1}{2} T^4 F_3(T)$, leading to
\be
\la q_{i}^2\ra_s^{\tiny\newmoon\newmoon} = \frac{\la \delta\tau\ra_s}{L} \frac{1}{6} \rho^2 T^2 F_3(T)  + {\cal O}(L^{-2}) \, .
\label{eq:correlee3c}
\ee
Putting all together, we obtain for the energy current noise correlator
\ben
\la \zeta_{i,s} \zeta_{\ell,s}\ra &=& \delta_{i,\ell} \frac{\la \delta\tau\ra_s}{L} \left[2\rho(1-\rho) T^2 F_2(T) + \frac{1}{6} \rho^2 T^2 F_3(T) \right] \nonumber \\
&+& {\cal O}(L^{-2}) \, .
\label{eq:correlee4}
\een

In the diffusive scaling limit introduced in Section \S\ref{s3a}, see Eqs. \eqref{eq:space}-\eqref{eq:time}, we expect the particle and energy current noises to scale in the same way as their corresponding currents, see Eq. \eqref{eq:currfields}, i.e.
\be
\xi_{i,s} \to \frac{\la \delta \tau\ra_s}{L^{2}} \xi(x,t) \, , \qquad \zeta_{i,s} \to \frac{\la \delta \tau\ra_s}{L^{2}} \zeta(x,t) \, ,
\label{eq:corrfields}
\ee
so the current fields in the diffusive scale can be splitted in a main part and a noise field
\be
j(x,t) = \bar{j}(x,t) + \xi(x,t) \, , \qquad q(x,t) = \bar{q}(x,t) + \zeta(x,t) \, .
\ee
The main fields $\bar{j}(x,t)$ and $\bar{q}(x,t)$ are given now by the Fick-Fourier's constitutive relations \eqref{eq:FickFourier}-\eqref{eq:Dmatrix}, namely
\be
\left(\begin{array}{l}
\bar{j}\\
\bar{q}
\end{array}\right)=-\mathbb{D}(\rho,T) \left(\begin{array}{l}
\partial_{x}\rho\\
\partial_{x}T
\end{array}\right) \, . \label{eq:FickFourier2}
\ee
In addition, by combining Eqs. \eqref{eq:correlpp3}, \eqref{eq:correlpe2} and \eqref{eq:correlee4} with the scaling in \eqref{eq:corrfields}, we obtain for the correlators of the noise fields in the diffusive scaling limit
\ben
\la \xi(x,t) \xi(x't')\ra &=& \frac{1}{L} M_{11}(\rho,T) \delta(x-x') \delta(t-t') \, , \nonumber \\
\la \xi(x,t) \zeta(x't')\ra &=& \frac{1}{L} M_{12}(\rho,T) \delta(x-x') \delta(t-t') \, , \label{eq:corrfields2} \\
\la \zeta(x,t) \zeta(x't')\ra &=& \frac{1}{L} M_{22}(\rho,T) \delta(x-x') \delta(t-t') \, , \nonumber 
\een
where we have taken into account that
\ben
L \delta_{i,\ell} = \frac{\delta_{i,\ell}}{\Delta x} &\to& \delta(x-x') \, , \label{eq.deltax} \\
\frac{L^3}{\la \delta\tau\ra_s} \delta_{s,s'} = \frac{\delta_{s,s'}}{\Delta t} &\to& \delta(t-t') \, , \label{eq.deltat}
\een
where $\Delta x\equiv 1/L$, and $\Delta t\equiv \la \delta\tau\ra_s/L^3$ is the time interval in the macroscopic diffusive scale corresponding to a typical microscopic time interval $\la \delta\tau\ra_s/L$, see discussion below Eq. \eqref{eq:coll}. The mobility transport coefficients in \eqref{eq:corrfields2} are defined as
\ben
M_{11}(\rho,T) &=& 2 \rho(1-\rho) F_0(T) \, , \nonumber \\
M_{12}(\rho,T) &=& 2 \rho(1-\rho) T F_1(T) = M_{21}(\rho,T) \, , \label{eq:mobil} \\
M_{22}(\rho,T) &=& 2 \rho(1-\rho) T^2 F_2(T) + \frac{1}{6} \rho^2 T^2 F_3(T) \, . \nonumber
\een
These transport coefficients are the elements of a \emph{symmetric} mobility matrix $\mathbb{M}(\rho,T)$ which controls the coupled fluctuations of the particle and energy current fields. Note that, as anticipated at the begining, the noises perturbing the current fields are weak, i.e. their correlators are inversely proportional to the system size in the $L\gg 1$ limit. 

Interestingly, the mobility matrix $\mathbb{M}(\rho,T)$ controlling the amplitude of current fluctuations in the KEP can be simply related to the Onsager's matrix $\mathbb{L}(\rho,T)$ of phenomenological transport coefficients in Eq. \eqref{eq:Onsager} associated with the dissipative fluxes, namely
\be
\mathbb{M}(\rho,T) = 2 \, \mathbb{L}(\rho,T) \, .
\label{eq:FDT2}
\ee
This is just an expression of the general fluctuation-dissipation theorem linking thermal fluctuations and the response to a small driving in microreversible systems (i.e. systems obeying detailed balance) \cite{de-groot13a}. This theorem can be easily demonstrated starting from the time reversibility of the dynamics, with an additional assumption on the Gaussian character of the associated fluctuations \cite{de-groot13a}. In this way, the validity of the fluctuation-dissipation theorem for the kinetic exclusion process, that we have just demonstrated, strongly supports the Gaussian character of the noise terms affecting the local current fields.

\section{Summary and conclusions}
\label{s5}

In this work we have derived the fluctuating hydrodynamics of a new model of transport, the kinetic exclusion process (KEP), characterized by the coupled nonlinear transport of two different locally-conserved fields. The kinetic exclusion process consists in energetic particles on a lattice subject to exclusion interactions, which jump and collide stochastically with energy-dependent rates. In addition the model can be coupled to different types of boundary baths, including both energy and particle\&energy reservoirs, which may drive the system out of equilibrium by introducing temperature and/or chemical potential gradients. Starting from the microscopic balance equations, we show that at a mesoscopic scale the KEP can be described by two coarse-grained fields, namely the particle density $\rho(x,t)$ and a temperature field $T(x,t)$, that evolve in time according to a pair of coupled continuity-like Langevin equations. The particle and energy current fields can be shown to be proportional to the gradients of the density and temperature fields using a local equilibrium approximation in the large system size limit, and we determine the transport coefficients which fully define these constitutive relations. These diffusivity transport coefficients are explicitly written in terms of the microscopic dynamics for the KEP, and are in general nonlinear functions of the local fields. Interestingly, the resulting hydrodynamic equations capture the possibility of (i) a particle flow in the absence of a density gradient, due exclusively to a temperature gradient (Soret effect), and (ii) an energy current due exclusively to the presence of a density gradient (Dufour effect) \cite{de-groot13a}. An analysis of the entropy production in the KEP allows us to identify the thermodynamic forces in the problem, from which we obtain the associated Onsager's matrix in terms of the diffusivity matrix. Moreover, we prove the positivity of entropy production in the KEP. We further analyze the hydrodynamic fluctuations which affect both particle and energy currents, deriving explicit expressions for the nonlinear amplitudes of the noise terms affecting both current fields, as well as their cross-correlations. This shows that these noises are weak, ${\cal O}(L^{-1})$, in the large system size limit ($L\gg 1$), and connects the mobility matrix measuring the amplitude of current fluctuations with the Onsager's matrix associated to the dissipative fluxes. We hence prove a fluctuation-dissipation theorem for the kinetic exclusion process, which supports the Gaussian character of current fluctuations as $L\to\infty$.

The kinetic exclusion process here introduced opens up new and exciting avenues of research in different directions. On one hand the KEP is a stochastic lattice gas characterized by two local conservation laws which are coupled nonlinearly, but it is still simple enough to be amenable to both analytical calculations and extensive computer simulations. This suggests to extend and generalize the formalism of macroscopic fluctuation theory \cite{bertini15a}, mostly applied up to now to simpler models with a single conservation law, to this more realistic setting, with the sight fixed on fully-hydrodynamic models of transport as in e.g. realistic fluids. In particular, we anticipate that the KEP will play an important role in the investigation of dynamical phase transitions and symmetry-breaking phenomena at the fluctuation level and how they are affected by multiple conservation laws. On the other hand, the kinetic exclusion process converges in two different limits to two key models of nonequilibrium physics, namely the simple symmetric exclusion process (SSEP) of diffusion \cite{derrida98a} and the Kipnis-Marchioro-Presutti (KMP) model of heat conduction \cite{kipnis82a}. These two models have been pivotal in the literature on exact results out of equilibrium, including the first rigorous derivation of the elusive Fourier's law of heat conduction from microscopic dynamics in the KMP model \cite{kipnis82a}, and one of the very few exact determinations of the steady state probability measure in a nonequilibrium system \cite{derrida98a,derrida01a}. These results, which should appear as limits of the KEP, suggest additional rigorous studies of the kinetic exclusion process. In particular, the existing mappings of both the KMP and SSEP models onto integrable quantum spin systems \cite{derrida98a,giardina07a,giardina09a} invite to seek a generalized spin mapping for the KEP which may offer new insights in the field of exactly-solvable models of nonequilibrium statistical physics.

\begin{acknowledgments}
We thank R. Hurtado for many useful comments on the manuscript. Financial support from Spanish Ministry MINECO project FIS2017-84256-P is also acknowledged. This study  has been partially financed by the \emph{Consejer\'{\i}a de Conocimiento, Investigaci\'on y Universidad} (Junta de Andaluc\'{\i}a), and European Regional Development Fund (ERDF), ref. SOMM17/6105/UGR.
\end{acknowledgments}

\bibliography{/Users/phurtado/Dropbox/PAPERS/BIBLIOGRAPHY/referencias-BibDesk-OK}{}


\vspace{1cm}

\onecolumngrid
\appendix

\section{Constitutive relation for the average energy current}
\label{app1}

In this appendix we compute the average energy current 
\be
\la q_{i}\ra_s = \frac{1}{L} \left\la \frac{\left[n_i\varepsilon_i -n_{i+1}\varepsilon_{i+1} + \frac{1}{2}(\varepsilon_{i+1}-\varepsilon_i)n_i n_{i+1}\right]~f(\varepsilon_i+\varepsilon_{i+1})}{\Omega_L(\bm{\nu})}\right\ra_s \, ,  
\label{aeq:curre2}
\ee
see Eq. (\ref{eq:curre2}) in the main text. To perform this average, we use the general expresion (\ref{eq:avegen4}), namely
\be
\frac{1}{L}\left\la \frac{g(\bm{\nu}_i,\bm{\nu}_{i+1})~f(\varepsilon_i+\varepsilon_{i+1})}{\Omega_L(\bm{\nu})} \right\ra_s = \frac{\la \delta\tau\ra_s}{L} \left\la g(\bm{\nu}_i,\bm{\nu}_{i+1}) f(\varepsilon_i+\varepsilon_{i+1}) \right\ra_s \, ,
\label{aeq:avegen3}
\ee
with $\la \delta\tau\ra_s = \lim_{L\to\infty} \la \Omega_{L}^{-1} \ra_s$ and
\be
\left\la g(\bm{\nu}_i,\bm{\nu}_{i+1}) f(\varepsilon_i+\varepsilon_{i+1}) \right\ra_s = \sum_{n_i,n_{i+1}=0,1} \int_0^\infty d\varepsilon_i d\varepsilon_{i+1}  ~g[(n_i,\varepsilon_i),(n_{i+1},\varepsilon_{i+1})] ~f(\varepsilon_i+\varepsilon_{i+1})~P_\text{LE}^{(i)}(n_i,\varepsilon_i;s) P_\text{LE}^{(i+1)}(n_{i+1},\varepsilon_{i+1};s) \, . 
\label{aeq:avegen4}
\ee
For the energy current, the local function to average is
\be
g(\bm{\nu}_i,\bm{\nu}_{i+1}) = n_i\varepsilon_i -n_{i+1}\varepsilon_{i+1} + \frac{1}{2}(\varepsilon_{i+1}-\varepsilon_i)n_i n_{i+1} \, .
\label{aeq:gfunc}
\ee
Splitting the local equilibrium probability measure for a given state of pair $(i,i+1)$ as was done in Eq. (\ref{eq:pLE0}) of the main text leads to
\be
\la q_i \ra_s =\la q_i \ra_s^{\tiny\newmoon\fullmoon} +  \la q_i \ra_s^{\tiny\fullmoon\newmoon} + \la q_i \ra_s^{\tiny\newmoon\newmoon} \, ,
\label{aeq:curre3}
\ee
where the different contributions are
\ben
\la q_i \ra_s^{\tiny\newmoon\fullmoon} &=&  \frac{\la \delta\tau\ra_s}{L} \int_0^\infty d\varepsilon_i d\varepsilon_{i+1}~\varepsilon_i f(\varepsilon_i +\varepsilon_{i+1})  P_{LE}^{\tiny\newmoon\fullmoon}(\varepsilon_{i},\varepsilon_{i+1};s) \, , \label{aeq:q10} \\
\la q_i \ra_s^{\tiny\fullmoon\newmoon} &=& - \frac{\la \delta\tau\ra_s}{L} \int_0^\infty d\varepsilon_i d\varepsilon_{i+1}~\varepsilon_{i+1} f(\varepsilon_i +\varepsilon_{i+1})  P_{LE}^{\tiny\fullmoon\newmoon}(\varepsilon_{i},\varepsilon_{i+1};s) \, , \label{aeq:q01} \\
\la q_i \ra_s^{\tiny\newmoon\newmoon} &=& - \frac{\la \delta\tau\ra_s}{2 L} \int_0^\infty d\varepsilon_i d\varepsilon_{i+1}~(\varepsilon_i-\varepsilon_{i+1}) f(\varepsilon_i +\varepsilon_{i+1})  P_{LE}^{\tiny\newmoon\newmoon}(\varepsilon_{i},\varepsilon_{i+1};s)  \, . \label{aeq:q11} 
\een
Using here the definitions (\ref{eq:pLE1})-(\ref{eq:pLE4}) for the particular local equilibrium probabilities, we find for the first two contributions
\ben
\la q_i \ra_s^{\tiny\newmoon\fullmoon} &=&  \frac{\la \delta\tau\ra_s}{L} \rho T \left(1-\rho-\frac{\partial_{x}\rho}{L}\right) F_1(T) \, , \label{aeq:q102} \\
\la q_i \ra_s^{\tiny\fullmoon\newmoon} &=& - \frac{\la \delta\tau\ra_s}{L} \rho \left(1-\rho\right) T \left[ \left(1+\frac{\partial_{x}\rho}{L\rho}-\frac{\partial_{x}T}{LT}\right) F_1(T) + \frac{\partial_{x}T}{LT} F_2(T) \right]\, , \label{aeq:q012}
\een
where we recall the definition of the $F_n(T)$ functions
\be
F_n(T) \equiv \int_0^\infty dy~\text{e}^{-y} y^n f(Ty) \, .
\label{aeq:Fn}
\ee
Finally, the average energy current due to particle collisions can be written as
\ben
\la q_i \ra_s^{\tiny\newmoon\newmoon} &=& - \frac{\la \delta\tau\ra_s}{2 L} \frac{\rho^2}{T^2} \left[\left(1+ \frac{\partial_{x}\rho}{L\rho} \right) \int_{0}^{\infty}d\varepsilon_{i} d\varepsilon_{i+1} \left(\varepsilon_{i}-\varepsilon_{i+1}\right) f\left(\varepsilon_{i}+\varepsilon_{i+1}\right) \text{e}^{-(\varepsilon_{i}+\varepsilon_{i+1})/T} \right. \nonumber \\ 
&-& \left. \frac{\partial_{x}T}{LT} \int_{0}^{\infty}d\varepsilon_{i} d\varepsilon_{i+1} \left(\varepsilon_{i}-\varepsilon_{i+1}\right)\left(1-\frac{\varepsilon_{i+1}}{T} \right) f\left(\varepsilon_{i}+\varepsilon_{i+1}\right) \text{e}^{-(\varepsilon_{i}+\varepsilon_{i+1})/T} \right] \, .
\label{aeq:q112}
\een
The first integral in the previous equation is clearly zero due to the antisymmetric character of the integrand under the exchange of the dummy integration variables, $\varepsilon_i \leftrightarrow \varepsilon_{i+1}$. On the other hand, the second integral (denoted now as $I_{\tiny\newmoon\newmoon}$) can be symmetrized by noting that the integral
\be
I_{\tiny\newmoon\newmoon}^* \equiv \int_{0}^{\infty}d\varepsilon_{i} d\varepsilon_{i+1} \left(\varepsilon_{i+1}-\varepsilon_{i}\right)\left(1-\frac{\varepsilon_{i}}{T} \right) f\left(\varepsilon_{i}+\varepsilon_{i+1}\right) \text{e}^{-(\varepsilon_{i}+\varepsilon_{i+1})/T} \, ,
\ee
is exactly equal to $I_{\tiny\newmoon\newmoon}$, as it is obtained from $I_{\tiny\newmoon\newmoon}$ by exchanging the dummy integration variables, $\varepsilon_i \leftrightarrow \varepsilon_{i+1}$. In this way $I_{\tiny\newmoon\newmoon} = \frac{1}{2}\left(I_{\tiny\newmoon\newmoon} + I_{\tiny\newmoon\newmoon}^* \right)$, and we obtain
\be
\la q_i \ra_s^{\tiny\newmoon\newmoon} = \frac{\la \delta\tau\ra_s}{4 L} \rho^2  \frac{\partial_{x}T}{LT^4} \int_{0}^{\infty}d\varepsilon_{i}d\varepsilon_{i+1} \left(\varepsilon_{i}-\varepsilon_{i+1}\right)^2  f\left(\varepsilon_{i}+\varepsilon_{i+1}\right) \text{e}^{-(\varepsilon_{i}+\varepsilon_{i+1})/T} \, .
\label{aeq:q113}
\ee 
This integral can be now computed in polar coordinates $(r,\phi)$ by changing variables so that $\sqrt{\varepsilon_{i}}=r\cos\phi$ and $\sqrt{\varepsilon_{i+1}}=r\sin\phi$, with $r\in [0,\infty)$ and $\phi\in[0,\pi/2]$. The Jacobian of this transformation is $J=4r^{3}\cos\phi \sin\phi$, and the average collision current simplifies to
\be
\la q_i \ra_s^{\tiny\newmoon\newmoon} = - \frac{\la \delta\tau\ra_s}{L^2} \frac{\rho^2}{T^4}  \partial_{x}T \underbrace{\left[\int_{0}^{\frac{\pi}{2}}d\phi\,\cos^{2}(2\phi) \cos\phi \sin\phi\right]}_{1/6}\int_{0}^{\infty}dr\,r^{7}f\left(r^{2}\right)\text{e}^{-r^2/T} = -  \frac{\la \delta\tau\ra_s}{L^2} \frac{\rho^2}{12} F_3(T) \partial_{x}T \, ,
\label{aeq:q114}
\ee
where we have changed variables to $y=r^2/T$ in the last equality, which leads to a $F_3(T)$ function, see Eq. (\ref{aeq:Fn}) above. Putting all terms together, see Eq. (\ref{aeq:curre3}), we find
\be
\la q_i \ra_s = \frac{\la \delta\tau\ra_s}{L^2} \left\{ -T F_1(T) \partial_x\rho - \left(\rho(1-\rho)[F_2(T)-F_1(T)] + \frac{\rho^2}{12} F_3(T) \right) \partial_x T\right\} \, .
\label{aeq:curre4}
\ee
This expression confirms again the heuristic scaling anticipated in the main text, see Eq. (\ref{eq:currfields}), and leads to the following constitutive relation for the energy current field in the diffusive scaling limit
\be
q(x,t) = -D_{21}(T) \partial_x\rho(x,t) - D_{22}(\rho,T) \partial_x T(x,t) \, , 
\label{aeq:curre5}
\ee
with the following transport coefficients
\ben
D_{21}(T) &=& T F_1(T)\, , \label{aeq:d21} \\
D_{22}(\rho,T) &=& \rho(1-\rho)[F_2(T)-F_1(T)] + \frac{\rho^2}{12} F_3(T) \, ,  \label{aeq:d22}
\een
Note that the constitutive relation (\ref{aeq:curre5}) captures the possibility of an energy flow in the absence of a temperature gradient, due exclusively to a density grandient. This is the well-known Dufour effect \cite{de-groot13a}.

\section{Integral for the collision contribution to the energy current correlator}
\label{app2}

In this appendix we compute the contribution of particle collisions to the energy current correlator. As explained in the main text, see Eq. \eqref{eq:correlee3c0} and the associated discussion, this contribution is captured by the following integral
\ben
\la q_{i}^2\ra_s^{\tiny\newmoon\newmoon} &=& \frac{\la \delta\tau\ra_s}{L} \int_0^\infty d\varepsilon_i d\varepsilon_{i+1} \int_0^1 d\alpha_s \left[(1-\alpha_s) \varepsilon_{i,s} - \alpha_s \varepsilon_{i+1,s} \right]^2 f(\varepsilon_i +\varepsilon_{i+1})  P_{LE}^{\tiny\newmoon\newmoon}(\varepsilon_{i},\varepsilon_{i+1};s) \nonumber \\
&=& \frac{\la \delta\tau\ra_s}{L} \frac{1}{3}\frac{\rho^2}{T^2} \int_0^\infty d\varepsilon_i d\varepsilon_{i+1} (\varepsilon_i^2 +\varepsilon_{i+1}^2 - \varepsilon_i \varepsilon_{i+1}) f(\varepsilon_i +\varepsilon_{i+1}) \text{e}^{-(\varepsilon_i +\varepsilon_{i+1})/T} + {\cal O}(L^{-2}) \, ,
\label{eq:correleeapp1}
\een
where we have used Eq. \eqref{eq:pLE1}, neglecting subdominant ${\cal O}(L^{-2})$ terms, together with the averages $\la \alpha_s^2\ra = 1/3 = \la (1-\alpha_s)^2\ra$ and $\la \alpha_s(1-\alpha_s)\ra = 1/6$. The last integral can be solved in simple terms using a change of variables to polar coordinates. In particular we now define, as in Appendix \ref{app1}, $\sqrt{\varepsilon_{i}}=r\cos\phi$ and $\sqrt{\varepsilon_{i+1}}=r\sin\phi$, with $r\in [0,\infty)$ and $\phi\in[0,\pi/2]$. The Jacobian of this transformation is $J=4r^{3}\cos\phi \sin\phi$, and the above integral transforms into
\ben
\la q_{i}^2\ra_s^{\tiny\newmoon\newmoon} &=& \frac{\la \delta\tau\ra_s}{L} \frac{4 \rho^2}{3 T^2} \int_{0}^{\frac{\pi}{2}}d\phi\, \cos\phi \sin\phi \left(\cos^4\phi + \sin^4\phi - \cos^2\phi \sin^2\phi \right)
\int_{0}^{\infty}dr\,r^{7}\text{e}^{-r^2/T} f\left(r^{2}\right) + {\cal O}(L^{-2})  \nonumber \\
&=& \frac{\la \delta\tau\ra_s}{L} \frac{4}{6} \rho^2 T^2  \underbrace{\left[\int_{0}^{\frac{\pi}{2}}d\phi\, \cos\phi \sin\phi \left(1-3 \cos^2\phi \sin^2\phi \right) \right]}_{1/4} \int_{0}^{\infty}dy\,y^{3}\text{e}^{-y} f\left(Ty\right) + {\cal O}(L^{-2}) \, ,
\label{eq:correleeapp2}
\een
where we have used that $\cos^4\phi + \sin^4\phi - \cos^2\phi \sin^2\phi = 1-3 \cos^2\phi \sin^2\phi$ in the angular integral, which can be now solved easily, as well as the change of variable $y=r^2/T$ in the radial integral, which now corresponds to the $F_3(T)$ function, see general definition in e.g. Eq. \eqref{aeq:Fn}. In this way, the integral of interest leads to the final result
\be
\la q_{i}^2\ra_s^{\tiny\newmoon\newmoon} = \frac{\la \delta\tau\ra_s}{L} \frac{1}{6} \rho^2 T^2 F_3(T)  + {\cal O}(L^{-2}) \, .
\label{eq:correleeap4}
\ee

\end{document}